\documentclass[trans]{IEEEtran}

\usepackage{xspace,amsmath,amssymb,amsfonts,epsfig,syntonly,amsthm}
\usepackage{cite,bm,color,url,textcomp}
\usepackage{algorithmicx,algorithm,algpseudocode}
\usepackage{epstopdf}
\usepackage{empheq}
\usepackage{balance}
\usepackage{diagbox}
\usepackage{makecell}
\usepackage{array}

\newcommand*{\red}{\color{black}}

\long\def\symbolfootnote[#1]#2{\begingroup
\def\thefootnote{\fnsymbol{footnote}}
\footnote[#1]{#2}\endgroup}

\makeatother
\psfull

\allowdisplaybreaks[3]

\begin{document}
\title{Modeling and Analysis of Energy Harvesting and  Smart Grid-Powered Wireless Communication Networks: A Contemporary Survey}

\author{Shuyan Hu, Xiaojing Chen, Wei Ni, Xin Wang, and Ekram Hossain
\thanks{
Work in this paper was supported by the National Natural Science Foundation of China under Grant 61671154, and the Innovation Program of Shanghai Municipal Science and Technology Commission under Grant 17510710400. The work of W. Ni was supported by the Fudan University Key Laboratory (State Key Lab of ASIC and System) Senior Visiting Scholarship.
The work of E. Hossain was supported in part by an ENGAGE Grant (EGP 533553-18) from the Natural Sciences and Engineering Research Council of Canada (NSERC).
The work of S. Hu and X. Chen was done when they were with the Department of Communication Science and Engineering, Fudan University.

S. Hu is with the State Key Laboratory of ASIC and System, the School of Information Science and Technology, Fudan University, Shanghai 200433, China
(e-mail: syhu14@fudan.edu.cn).

X. Chen is with the Shanghai Institute for Advanced Communication and Data Science, Shanghai University, Shanghai 200444, China
(e-mail: jodiechen@shu.edu.cn).

W. Ni is with the Commonwealth Scientific and Industrial Research Organization (CSIRO), Sydney, NSW 2122, Australia
(e-mail: wei.ni@data61.csiro.au).

X. Wang is with the State Key Laboratory of ASIC and System, the Shanghai Institute for Advanced Communication and Data Science, the Department of Communication Science and Engineering, Fudan University, Shanghai 200433, China
(e-mail: xwang11@fudan.edu.cn).

E. Hossain is with the Department of Electrical and Computer Engineering, University of Manitoba, Winnipeg, MB R3T 5V6, Canada
(e-mail: ekram.hossain@umanitoba.ca).

The first two authors contributed equally to this work.}
}
\maketitle

\setcounter{page}{1}

\begin{abstract}
The advancements in smart power grid and the advocation of ``green communications'' have inspired the wireless communication networks
to harness energy from ambient environments and operate in an energy-efficient manner for economic and ecological benefits.
This article presents a contemporary review of recent breakthroughs on the utilization, redistribution,
trading and planning of energy harvested in future wireless networks interoperating with smart grids.
This article starts with classical models of renewable energy harvesting technologies.
We embark on constrained operation and optimization of different energy harvesting wireless systems,
such as point-to-point, multipoint-to-point, multipoint-to-multipoint, multi-hop, and multi-cell systems.
We also review wireless power and information transfer technologies which provide a special implementation
of energy harvesting wireless communications. A significant part of the article is devoted to the redistribution of redundant (unused) energy harvested within cellular networks,
the energy planning under dynamic pricing when smart grids are in place,
and two-way energy trading between cellular networks and smart grids.
Applications of different optimization tools,
such as convex optimization, Lagrangian dual-based method,
subgradient method, and Lyapunov-based online optimization, are compared.
This article also collates the potential applications of energy harvesting techniques in emerging (or upcoming) 5G/B5G communication systems.
It is revealed that an effective redistribution and two-way trading of energy can significantly reduce the electricity bills of wireless service providers and decrease the consumption of brown energy.
A list of interesting research directions are provided, requiring further investigation.

\end{abstract}


\section{Introduction}
Future mobile cellular communication networks are envisaged to be rolled out with a dramatically increasing number of cells and continuously reducing cell size, due to explosive mobile traffic~\cite{xiaojing2017}.
The traffic volume in the emerging fifth-generation (5G) systems and future systems beyond 5G (B5G) is estimated to be tens of Exabytes per month, expecting the capacity of 5G/B5G networks to be $1000$ times higher than that of current cellular networks \cite{1000x, buzzi}.
The thousand-fold increase of system capacity must be achieved with a similar or even lower power level than today's \cite{white, andrews}.
Increasing the network energy efficiency (EE) has been pursued by the GreenTouch consortium \cite{greentouch, foundation}.
Huawei has also deployed solar-powered base stations (BSs) in Bangladesh \cite{huawei}.
Ericson and Nokia Siemens Networks have designed green BSs with renewable power supplies,
such as wind turbines and solar panels, to reduce the consumption of fuel generated electricity \cite{eric, nokia}.
Energy-efficient techniques, such as BS switching~\cite{bi2019joint}, offline power allocation~\cite{inan2016online}, and online data scheduling~\cite{Chen2016Optimal,chen2015provisioning}
have been developed to reduce power consumption or increase network capacity.

Along with the development of cellular networks, power grid is also undergoing a radical revolution. Rapidly emerging smart grids, enabled with smart meters, are expected to provide new intelligent functionalities, e.g., decentralized power production/generation, bidirectional (also known as ``two-way'') energy trading, energy redistribution, and request management/coordination~\cite{fang2012smart}.
Cellular networks, as integrating components (or elements) of smart grids, can support effective energy utilization and redistribution, and price negotiation by interoperating with smart grids~\cite{Xu15, xiaolu}.

Smart grid and ``green communications'' have spawned extensive studies recently.
Several magazine articles \cite{chen2015provisioning, Xu15, mao215, xiaojing2017, Gunduz14design, Ozel15funda}
and survey papers \cite{buzzi, qq17, alsa18, ehcomm, ehrf, ehsensor, pras, perera18, erol15, rehmani19 }
review energy-efficient and energy harvesting (EH) powered communication networks from a range of different angles.
However, none of the existing reviews captures comprehensively the constrained wireless operations powered by renewable energy sources (RES) and their underlaying optimization methodologies, and the interoperability between wireless networks and smart grids (as done in this article).
Energy-efficient techniques for 5G networks
are summarized in \cite{buzzi} and \cite{qq17}, from the aspects of network deployment, energy/spectrum efficiency,
and delay/bandwidth versus power.
Featuring energy-harvesting wireless networks, beamforming techniques are reviewed in~\cite{alsa18}.
Energy scheduling, optimization, and application are presented in \cite{ehcomm}.
Circuit design, hardware implementation, EH techniques, and communication protocols
are reviewed in~\cite{ehrf, ehsensor, pras, perera18} with specific emphases on radio frequency networks~\cite{ehrf},
sensor networks \cite{ehsensor, pras}, and joint information and power transfer systems \cite{perera18}.
Smart grid and its merits in improving the operations of wireless networks
and software-defined networks are reported in \cite{erol15} and \cite{rehmani19}, respectively.

\begin{figure}
\centering
\includegraphics[width=0.47\textwidth]{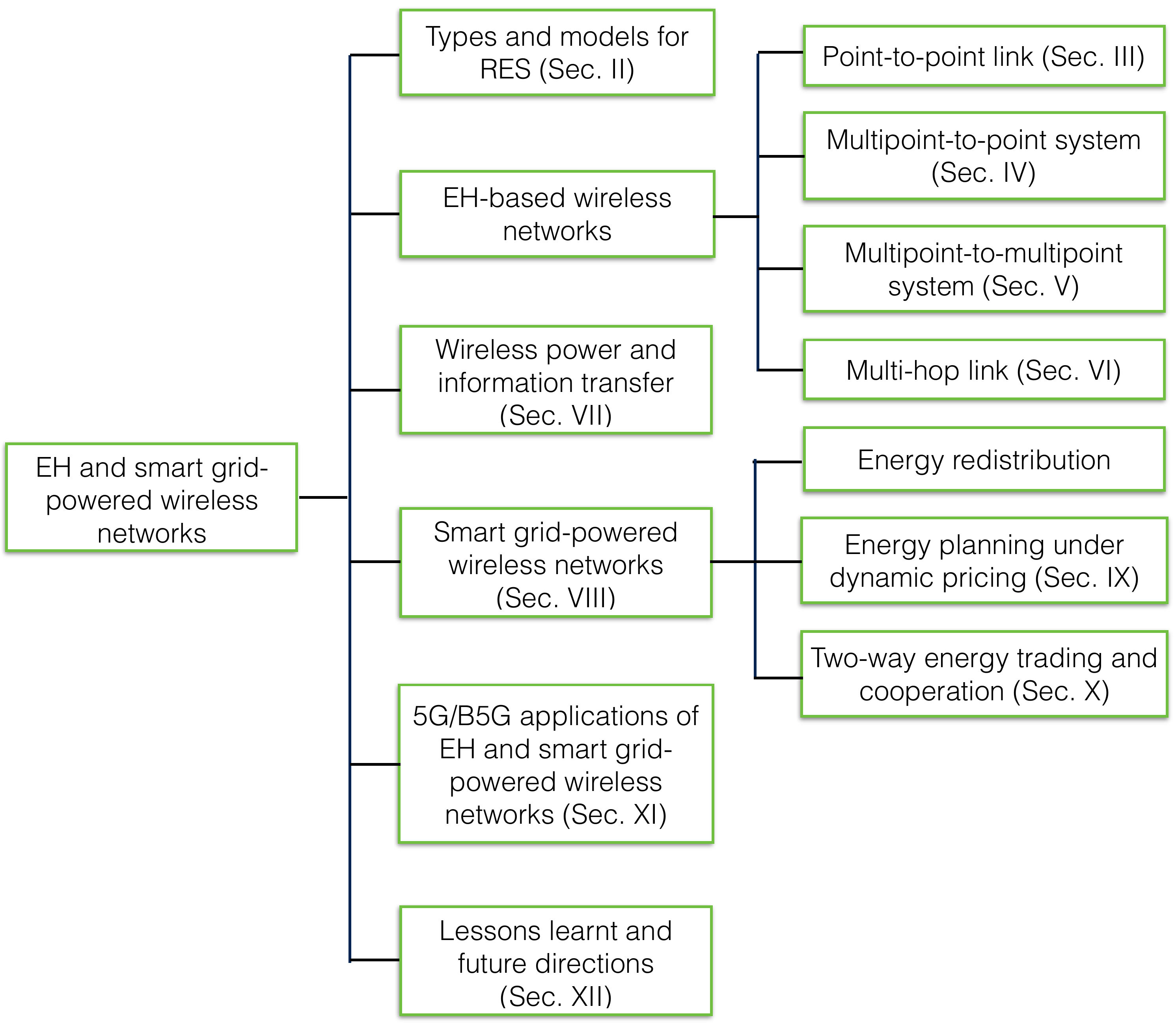}
\caption{The main structure of this article.}
\label{diagram}
\end{figure}

{\em This article takes a new and unique angle of the interoperability of wireless communication networks and smart grids.
Recent breakthroughs  on the utilization, redistribution, trading and planning of energy harvested
in future wireless communication networks interoperating with smart grids are reviewed}.
We start with state-of-the-art models of renewable EH technologies.
Then, we embark on constrained wireless operations
subject to non-persistent (renewable) energy harvested from ambient environments.
A range of energy-harvesting wireless communication systems, such as single point-to-point, multipoint-to-point,
multipoint-to-multipoint, multi-hop, and multi-cell systems.
We also go through a special yet widely studied class of EH techniques, where radio signals also play the role to deliver energy and joint
optimizations of wireless power and information delivery are carried out.

Performance metrics of energy-aware/constrained wireless operations include energy and operational cost minimization,
and utility maximization, by considering power allocation, transmit beamforming, traffic load, and users' quality-of-service (QoS).
Many classical techniques and methodologies are applied, such as convex optimization, Lagrangian dual-based method, game theory,
dynamic programming (DP), (stochastic) subgradient method (SGM), Lyapunov-based online optimization, and so forth.
Different system models and optimization criteria allow us to characterize, quantify, design and compare
different operating strategies of wireless networks from different perspectives, when practical design parameters arise.

A substantial part of this article is devoted to the latest techniques on the redistribution
of redundant energy within cellular networks, energy planning under dynamic pricing,
and bidirectional energy trading of cellular networks, as well as smart grids.
Empowered by renewable energy sources (RES), wireless power transfer (WPT), and/or smart grid,
actions of energy management such as the harvesting of RES, and the redistribution and (predictive) trading of redundant energy
can be performed in the wireless networks to achieve green and energy-efficient operations.
As will be revealed, the use of RES can significantly reduce the electricity bills of wireless service providers
and decrease the consumption of brown energy, i.e. coal and oil.
The article also discusses the potential applications of EH in future 5G/B5G networks,
and emerging technologies such as mobile edge computing (MEC), deep learning, ultra reliable and low latency communication (URLLC), non-orthogonal multiple access (NOMA), and so on.
As will be discussed, the energy management from the demand side can have a profound impact on future wireless networks.

By taking the new angle of the interoperability of wireless communication networks and smart grids
(as compared to the existing surveys~\cite{buzzi, qq17, alsa18, ehcomm, ehrf, ehsensor, pras, perera18, erol15, rehmani19}),
we organize the rest of this article in the following way.
Different types of RES and their popular mathematical models are colated in Section \ref{sec.res},
which are salient for design optimization of different types of RES-powered wireless links and networks in
Sections \ref{sec.single} to \ref{sec.multihop}.
In Section \ref{sec.wpt}, the joint wireless power and information transfer system is reviewed to provide a different perspective
of how energy and information signals interact in wireless communication networks.
From Sections \ref{sec.sharing} to \ref{sec.trading}, we discuss the roles of wireless networks as energy consumer,
harvester, and generator in the context of smart grid,
and thus the interoperability between wireless networks and smart power grids.
Specifically, we investigate the redistribution of redundant energy in wireless networks,
energy planning under dynamic pricing policy of the smart grid,
and bidirectional energy trading of EH wireless networks and smart grids.
In Section \ref{sec.app}, exciting 5G applications of EH, RES,
and smart grid-powered wireless devices are discussed.
In Section \ref{sec.future}, potential research opportunities are summarized, followed by concluding remarks in Section XIII.
Fig. \ref{diagram} provides the diagram to show the organization of this survey.

\section{Renewable Energy Sources (RES)}\label{sec.res}

\subsection{Types of Renewable Energy}
EH (or scavenging) is a
series of actions to collect and transform environmental renewable energy into electrical energy.
Unlike coal and oil, renewable energy can be regenerated with a wide range of different methods.
There are various types of RES, such as solar energy \cite{ehsolar, ehsolar2},
wind energy \cite{ehwind, ehwind2, ehtransmission}, electromagnetic (EM) radiation energy \cite{ehcomm, ehnear, ehrf, ehfar},
thermoelectric energy \cite{ehthermal, ehthermo, ehhuman}, and biomass energy \cite{ehbiomass}.

\subsubsection{Solar Energy}
One of the most
favored environmental RES is solar energy which has been extensively applied to different scenarios and applications
\cite{ehsolar, ehsolar2, ehsolarapp1, ehsolarapp2}.
Sunlight radiation is transformed into electric power via photovoltaic cells,
and then serves as the energy supply for self-supportable devices.
Although a potentially inexhaustible amount of energy can be obtained in this way, the energy usable to a device
could vary drastically even within a short duration in reality.
Furthermore, the EH level can be swayed by multiple factors, for instance, time of the day, solar elevation angle, weather conditions,
environmental prerequisites, and characteristics of photovoltaic cells.
These factors render the solar energy to be uncontrollable, unpredictable and nondispatchable.
Typically, the amount of solar energy is in the order of $100$ mW/cm$^2$~\cite{ehcomm}.

\subsubsection{Wind Energy}
Wind energy can be extracted by capturing the motion of the wind by a wind turbine.
The rotor speed output is used to perform maximal power point tracking.
The rotor frequency data is sent to a frequency-to-voltage (FV) converter which produces an appropriate voltage signal.
Wind energy can be also obtained by utilizing the movement of an anemometer lever to activate an alternator
and then using a pulsed buck-boost converter to transform the movement to electric power~\cite{ehsensor}.
Typically, when the rotor diameter is $1$ m and the wind speed is $8$ m/s,
the amount of energy that can be harvested is around $85$ W \cite{mao215}.

\subsubsection{Electromagnetic (EM) Radiation Energy}
Harvesting energy from EM radiation has
provided convenient power supplies for networks.
Featuring short-distance or long-distance scenarios,
the types of EM power supplies can be categorized into two groups: near-field and far-field EM radiation energy.
In the near-field scenarios, EM induction and magnetic resonance approaches~\cite{ehnear2, ehnear3}
often produce electric power within the range of a wavelength, and hence,
the power transfer efficiency can exceed 80$\%$ in the near-field scenarios \cite{ehnear}.
In the far-field scenarios up to several kilometers, the EM radiation, propagating in the fashion of radio frequency (RF) or microwave signals,
is received by antennas and transformed to electricity by rectifier circuits~\cite{ehrf, ehfar}.
The RF/microwave signals could come from beamforming signals sent by a given transmitter
or environmental EM radiations from the vicinity~\cite{fang15}.
Although the power density at the receiving end is related to the energy level of practical suppliers and the EM wave transmission distance,
this type of RES can be readily utilized, managed, and forecasted,
irrespective of time, location, and weather condition.

\subsubsection{Thermoelectric Energy}
The thermoelectric effect can be
utilized for EH.
In particular, a voltage signal can be generated between two conductors made of different materials
when their intersections are placed under different temperatures.
In practice, such a temperature gradient can take place in human bodies or machine operations.
The power densities of thermoelectric generations depend typically on the properties and temperature
difference of materials.
Their values are comparatively low and lie in the extent from $10$ $\mu$W/cm$^2$ to $1$~mW/cm$^2$~\cite{ehcomm}.

\subsubsection{Biomass Energy}
Since the aforementioned methods are not applicable for underwater EH,
bacterial metabolic activities have been exploited by microbial fuel cells (MFCs) to generate electricity
directly from broken down substratum.
Natural water contains abundant varieties of microorganisms and nutrients which are ideal for underwater EH by MFCs.
The amount of the energy is generally $153$ mW/cm$^2$ \cite{mao215}.

\subsection{Mathematical Models for Renewable Energy}
\subsubsection{Uncertainty Sets}
The a-priori knowledge of the stochastic RES amount $E_i^t$ is hardly available.
Yet, they can be potentially inferred and estimated from historical data.
Such estimations are generally bounded by uncertainty sets which characterize the ranges of the forecasted RES amounts.
To account for the temporally correlated RES amounts, two uncertainty sets ${\cal E}_i (i=1, 2, ...)$ are proposed
from the prospect of computational malleability.

The first model is given by a polyhedral set \cite{Zha13}:
\begin{align}
{\cal E}_i^{\textrm{p}} := \Bigg\{\mathbf{e}_i\;|\; &\underline{E}_i^t \leq E_i^t \leq \overline{E}_i^t, \notag \\
&\hspace{-1cm} E_{i,s}^{\min} \leq \sum_{t \in {\cal T}_{i,s}} E_i^t \leq E_{i,s}^{\max}, \; {\cal T}=\bigcup_{s=1}^S {\cal T}_{i,s}\Bigg\} \label{eq.Ei}
\end{align}
where $\overline{E}_i^t$ (or $\underline{E}_i^t$ ) is the maximum (or minimum) value of $E_i^t$;
and the operating period ${\cal T}$ is divided into non-overlapping, adjacent, and much smaller regions ${\cal T}_{i,s}$, $s = 1, \ldots, S$.
Each of the smaller regions can include several time slots. The overall amount of energy harvested at BS $i$ over the $s$-th smaller region
is constrained
by $E_{i,s}^{\min}$ and $E_{i,s}^{\max}$.

The second model amounts to
an uncertainty solution region with an ellipsoidal shape~\cite{ChenDG13}
\begin{align}
{\cal E}_i^{\textrm{e}} := \left\{\mathbf{e}_i = \hat{\mathbf{e}}_i + \bm{\varsigma}_i \;
|\; \bm{\varsigma}_i'\bm{\Sigma}^{-1}\bm{\varsigma}_i \leq 1\right\} \label{eq.Ei2}
\end{align}
where $\hat{\mathbf{e}}_i := [\hat{E}_i^1 ,\ldots, \hat{E}_i^T]'$ denotes the nominal EH amount at the $i$-th BS, and provides the predictable energy level; in other words, the expected (or mean) energy level.
$\bm{\varsigma}_i$ is a vector collecting the forecast errors.
The given matrix $\bm{\Sigma}\succ \mathbf{0}$ depicts the outline of the ellipsoid ${\cal E}_i^{\textrm{e}}$,
and thus decides the accuracy of the prediction.

In an attempt to account for random realizations of RES productions,
an uncertainty set can be in any form.
Based on the first- and second-order statistics of the stochastic amounts,
polyhedral or ellipsoidal sets are the most popular resorts \cite{dimi11}.
These two kinds of sets can be empirically obtained from historical statistics, and often used to help solve the relevant optimization problems.
Therefore, they can be implemented to EH systems as well.

\subsubsection{Stationary Process}
In some cases, the stochastic RES amounts are considered as
an independently and identically distributed (i.i.d.) process.
Weibull and Beta distributions are fairly accurate in portraying the traits of wind speed oscillation and solar power alteration, respectively~\cite{yeh08, wen15, abdu}.
The random RES amount can also be approximated by a Gaussian generation process \cite{ghaz17}.

\subsubsection{Markov Chain}
Different from the i.i.d. process, the Markov process implies that
the probability distribution of the $n$-th random variable is a function of that of the previous random variable in the process.
For RES-powered wireless sensor networks, the Markov chains are proposed to model the EH
process under the assumption that the time is slotted, a $2N$-state model can be designed
to represent the EH state (``active'' or ``inactive'') with a probability and the residual energy in the battery \cite{mkv2n}.
A state transition takes place when energy is harvested.
Under the same assumption of slotted time, a discrete-time Markov chain is leveraged to quantify the energy charged into a battery \cite{mkvdis}.
Assuming that energy can be replenished through EH and/or replacement of the battery,
a continuous-time Markov chain is designed to model the state of the battery in \cite{mkvcontinue}.
The state transitions are described as the different rates of EH and battery replacement,
each of which follows an independent Poisson process.
A stationary Markov process is used to model solar EH behavior, while a non-stationary one is
more effective in capturing vibrations introduced by other types of RES \cite{mkveh, mkvsolar}.

\section{Energy Harvesting-Based Communication over Point-to-Point Wireless Link}\label{sec.single}

\begin{figure*}
\centering
\includegraphics[width=1.03\textwidth]{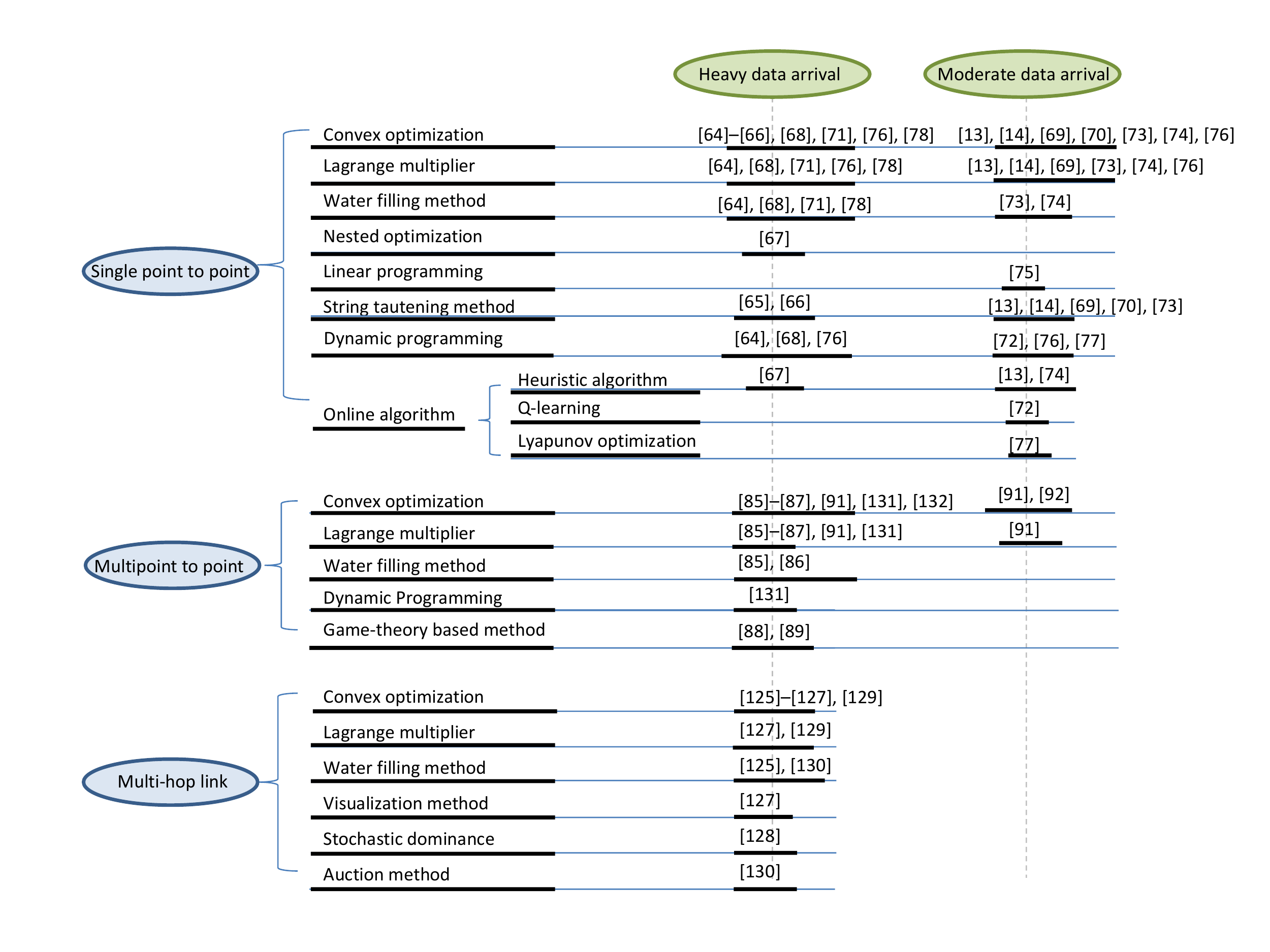}
\caption{A summary of optimization methods for resource allocation in EH-based point-to-point,
multipoint-to-point, and multi-hop wireless links.}
\label{roadmap}
\end{figure*}

There are many recent research works on powering data transmission over point-to-point radio link,
where the transmitters are typically sensors deployed remotely with no access to power grid
and harvesting RES from ambient environments all year around.
The sensors can be used to monitor temperature~\cite{Hou2018Thermal}, rainfall~\cite{Kar2016On},
bush fire~\cite{Somov2014Circuit}, and wildlife~\cite{Karim2012Reliable}.
The key techniques and methodologies applied are convex programming,
Lagrange multiplier method, and dynamic programming (DP), as summarized in Fig.~\ref{roadmap}.
Convex problems can be solved by off-the-shelf general-purpose convex solvers. However, general-purpose
convex programming solvers require high-order multiplications and many iterations, leading to a high-order polynomial complexity and slow convergence \cite{boyd}. Also, the general-purpose solvers cannot unveil the underlying structure of the optimal data transmission policy. To this end, the Lagrange multiplier method (typically in coupling with the celebrated Karush-Kuhn-Tucker (KKT) optimality conditions \cite{boyd}) has been widely applied in the literature for simpler and more insightful solutions \cite{boyd}.
By leveraging the Lagrange multiplier method along with the KKT optimality conditions to convex problems, one can find
the global optimum of the problems subject to the inequality constraints.
DP generally simplifies a sophisticated optimization problem by decoupling it to be several subproblems and solving the subproblems recursively \cite{Bertsekas}. However, DP is subjected to the curse of dimensionality.

The arrival processes of data traffic modeled by existing works can be categorized to two types: heavy data arrival scenario \cite{Ozel2011Transmission} and moderate data arrival scenario \cite{Chen2016Optimal}. The former type assumes that the data to be transmitted arrive before the start of the transmissions \cite{Ozel2011Transmission}; in other words, there are always data available in the transmit buffer. The latter type, under a more general assumption, considers that the data arrive during the course of transmission \cite{Chen2016Optimal}.

\subsection{Heavy Data Arrival}

\begin{table*}[t]
\renewcommand{\arraystretch}{1.5}
\centering
\caption{A summary of optimization problems for EH point-to-point links}\label{tab:algo}
    \begin{tabular}{ | c | c | c | c |}
    \hline
     \diagbox{\makecell{Settings}}{\makecell{Objectives}}

    &\makecell{Throughput maximization }    &\makecell{EE maximization}     &\makecell{Transmission completion time minimization}  \\ \hline

Time-invariant channel&\cite{Tutuncuoglu2012Optimum,Bai2011Throughput,Xu2012Throughput,yuan15}&\cite{Xu2012Throughput,Chen2016Optimal,chen2014Globecom,chen2015provisioning}& \cite{Tutuncuoglu2012Optimum,Jing2012Optimal}           \\ \hline
Time-varying channel&\cite{Ozel2011Transmission,Orhan2012Throughput,yuan15,blas}&\cite{Chen2014ICC,gong13,kang14,ahmed13,cui15}&    \cite{Ozel2011Transmission}           \\ \hline
Discrete energy arrivals &\cite{Tutuncuoglu2012Optimum,Ozel2011Transmission,Orhan2012Throughput,yuan15,blas}&\cite{Chen2016Optimal,Chen2014ICC,chen2014Globecom,chen2015provisioning,gong13,kang14,ahmed13,cui15}&    \cite{Tutuncuoglu2012Optimum,Ozel2011Transmission,Jing2012Optimal}          \\ \hline
Continuous energy arrivals &\cite{Bai2011Throughput,Xu2012Throughput}&\cite{Xu2012Throughput}& --            \\ \hline
Finite battery size&\cite{Tutuncuoglu2012Optimum,Ozel2011Transmission,Bai2011Throughput,Orhan2012Throughput,yuan15,blas}&\cite{chen2014Globecom,chen2015provisioning,gong13,ahmed13}&  \cite{Tutuncuoglu2012Optimum,Ozel2011Transmission}            \\ \hline
Strict Deadline &\cite{Tutuncuoglu2012Optimum,blas}&\cite{Chen2014ICC,Chen2016Optimal,chen2014Globecom,chen2015provisioning}&   \cite{Tutuncuoglu2012Optimum}          \\ \hline
Circuit power consumption &\cite{Bai2011Throughput,Orhan2012Throughput,Xu2012Throughput}&\cite{Xu2012Throughput,Chen2016Optimal,chen2015provisioning}& --              \\ \hline
 Offline  Algorithm           & \cite{Tutuncuoglu2012Optimum,Ozel2011Transmission,Bai2011Throughput,Orhan2012Throughput,Xu2012Throughput,yuan15,blas}       &  \cite{Xu2012Throughput,Chen2016Optimal,Chen2014ICC,chen2014Globecom,chen2015provisioning,gong13,kang14,ahmed13}    &     \cite{Tutuncuoglu2012Optimum,Ozel2011Transmission,Jing2012Optimal}        \\ \hline
Online Algorithm & \cite{Ozel2011Transmission,Xu2012Throughput,yuan15,blas}  &\cite{Xu2012Throughput,Chen2016Optimal,gong13,kang14,ahmed13,cui15}&        \cite{Ozel2011Transmission}     \\ \hline
  \end{tabular}
\end{table*}

By assuming a saturated traffic condition, the works in \cite{Tutuncuoglu2012Optimum,Ozel2011Transmission,Bai2011Throughput,Orhan2012Throughput,Xu2012Throughput,yuan15} focus on the impact of EH on the optimal transmit power. The works aim to design a transmission policy specifying the transmit power over the interval $[0, T]$ to maximize the overall throughput. Given a total amount of data to send, the minimization of the completion time is also investigated \cite{Tutuncuoglu2012Optimum,Ozel2011Transmission}.

\subsubsection{Idle Circuit Power}
A visualization method is used in \cite{Tutuncuoglu2012Optimum} to unveil the optimal transmission policy. Let $E_t$ denote the amount of energy scavenged at time $t$. Let $\tilde{E}_t$ denote the cumulative curve of harvested energy, i.e., the total amount of energy harvested by time $t$ (or $\tilde{E}_t=\sum_t E_t$). Similarly, let $\tilde{P}_t$ denote the cumulative curve of transmitted energy, i.e., $\tilde{P}_t =\sum_t P_t$. To determine a transmission schedule policy is to specify the non-decreasing and continuous function $\tilde{P}_t$.

\begin{figure}
\centering
\includegraphics[width=0.45\textwidth]{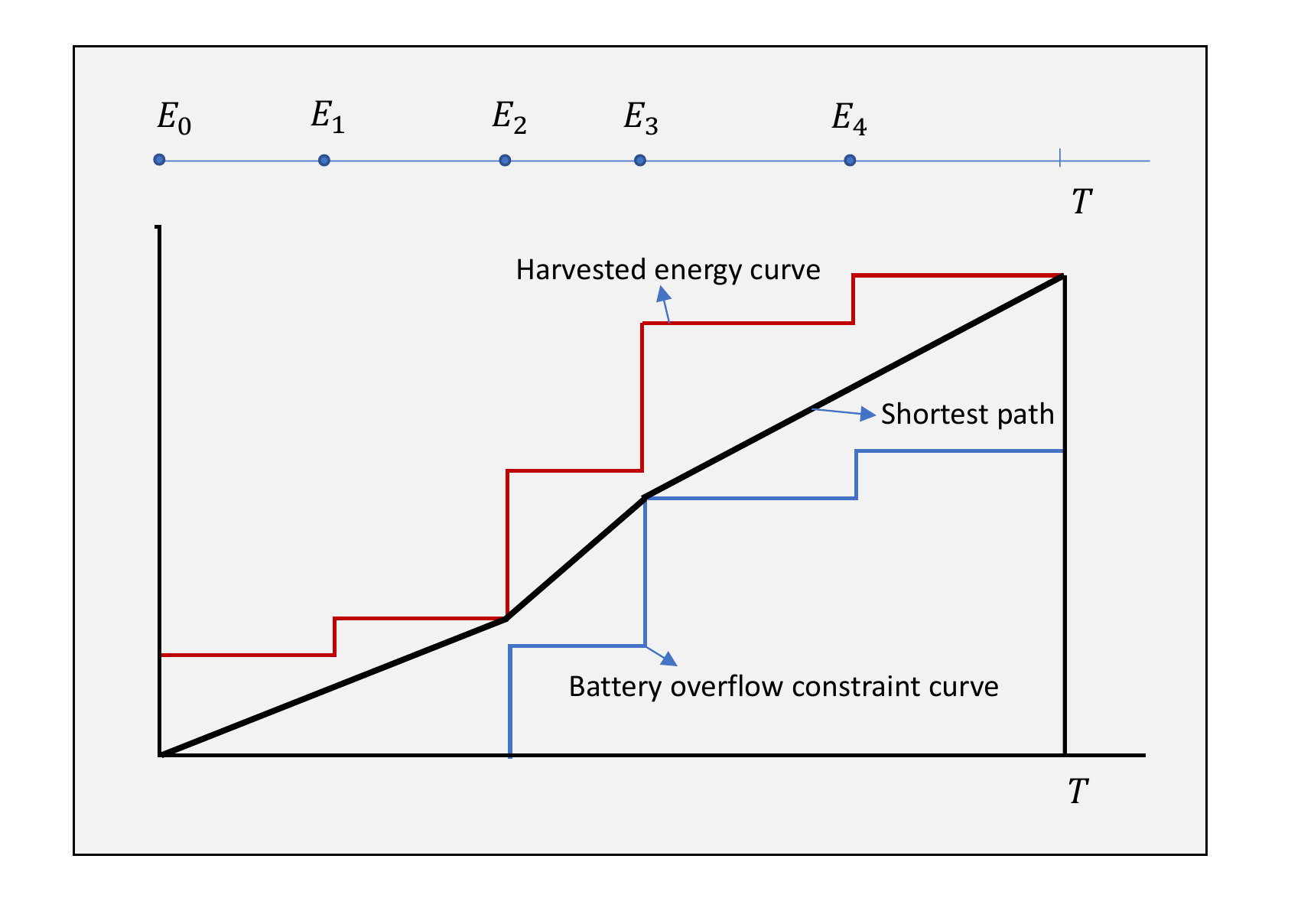}
\caption{Optimal transmission policy is the shortest path in the energy feasibility region \cite{Ozel15funda}.}
\label{curve}
\end{figure}

An inherent constraint is \textit{energy causality}, i.e., no energy can be utilized before it is harvested \cite{Gunduz14design,Ozel15funda}. This indicates that the curve of transmitted energy $\tilde{P}_t$ should lie under the curve of harvested energy $\tilde{E}_t$ all the time. With finite battery capacity $E_{\max}$, the optimal transmission policy should prevent the battery from being overcharged, i.e., the gap between the curves of transmitted energy and harvested energy (i.e. $\tilde{E}_t-\tilde{P}_t$) should not exceed the battery capacity. This constraint is referred to as \textit{no-energy-overflow} \cite{Ozel15funda}. Last but not the least, to deliver the maximum overall throughput or the minimum response time during the scheduling procedure, all the available energy is expected to be consumed by the deadline. In other words, the curve of transmitted energy $\tilde{P}_t$ needs to meet the curve of harvested energy $\tilde{E}_t$ at time instant $T$.
Collectively, these three constraints reveal that the optimal curve of transmitted energy $\tilde{P}_t$ starts from the origin, ends at point $(T, \tilde{E}_t)$, and lies between $\tilde{E}_t$ and $\tilde{E}_t-E_{\max}$; see Fig.~\ref{curve}. Here, the intersection of $\tilde{P}_t$ and $\tilde{E}_t$ means that the transmitter runs out of energy (or the battery is empty), and the intersection of $\tilde{P}_t$ and $\tilde{E}_t-E_{\max}$ denotes that the battery is full at the instant.

With the convexity of the transmit power $P(r)$ in regards to the transmit rate $r$, a use of Jensen's inequality \cite{boyd} can prove that employing a constant power can maximize the total volume of data delivered before its due time. In other words, the optimal strategy can be obtained by keeping a consistent power subject to the energy feasibility constraints. It is subsequently unveiled in \cite{Tutuncuoglu2012Optimum} by analyzing the original problem of interest and its constraints that the optimal transmission schedule which maximizes the throughput obeys the following two rules:
\\

\textit{{\bf Rule 1:} The transmit power only changes at the instants when the battery is completely drained or fully charged.\cite{Tutuncuoglu2012Optimum}.}

\textit{{\bf Rule 2:} The transmit power increases only at the instants when the battery is completely drained, and decreases only at the instants when the battery is fully charged \cite{Tutuncuoglu2012Optimum}.}
\\

Following Rules 1 and 2, the optimal transmission policy in~\cite{Tutuncuoglu2012Optimum} is shown to be the shortest path between the origin and end points in Fig.~\ref{curve}. It can be obtained by tautening a string between the harvested energy curve $\tilde{E}_t$ and the battery overflow constraint curve $\tilde{E}_t-E_{\max}$. Such a way of producing the optimal transmission policy is therefore called the ``String Tautening'' method \cite{Chen2016Optimal,chen2015provisioning}.
The optimal transmission policy in \cite{Tutuncuoglu2012Optimum} is developed over time-invariant channels.

For practical time-varying channels, Ozel et al. \cite{Ozel2011Transmission} propose a directional water-filling algorithm under energy causality and battery overflow constraints.
By formulating a convex problem maximizing the total throughput, the Lagrange multiplier method and the KKT optimality conditions are exploited to achieve the optimal solution. It can be concluded from Fig.~\ref{roadmap} that the majority of the water-filling methods in existing works are derived through convex formulation and Lagrange multipier method. By applying the Lagrange multiplier method to convex problems, one can find the globally optimal solution of the problems subject to the equality constraints, without explicit parameterization in terms of the constraints. The KKT conditions generalize the Lagrange multiplier method for solving inequality constrained optimization problems. In~\cite{Tutuncuoglu2012Optimum}, the KKT optimality conditions are used to deal with the inequality constraints accounting for the causality of EH and battery overcharging.

\begin{figure}
\centering
\includegraphics[width=0.44\textwidth]{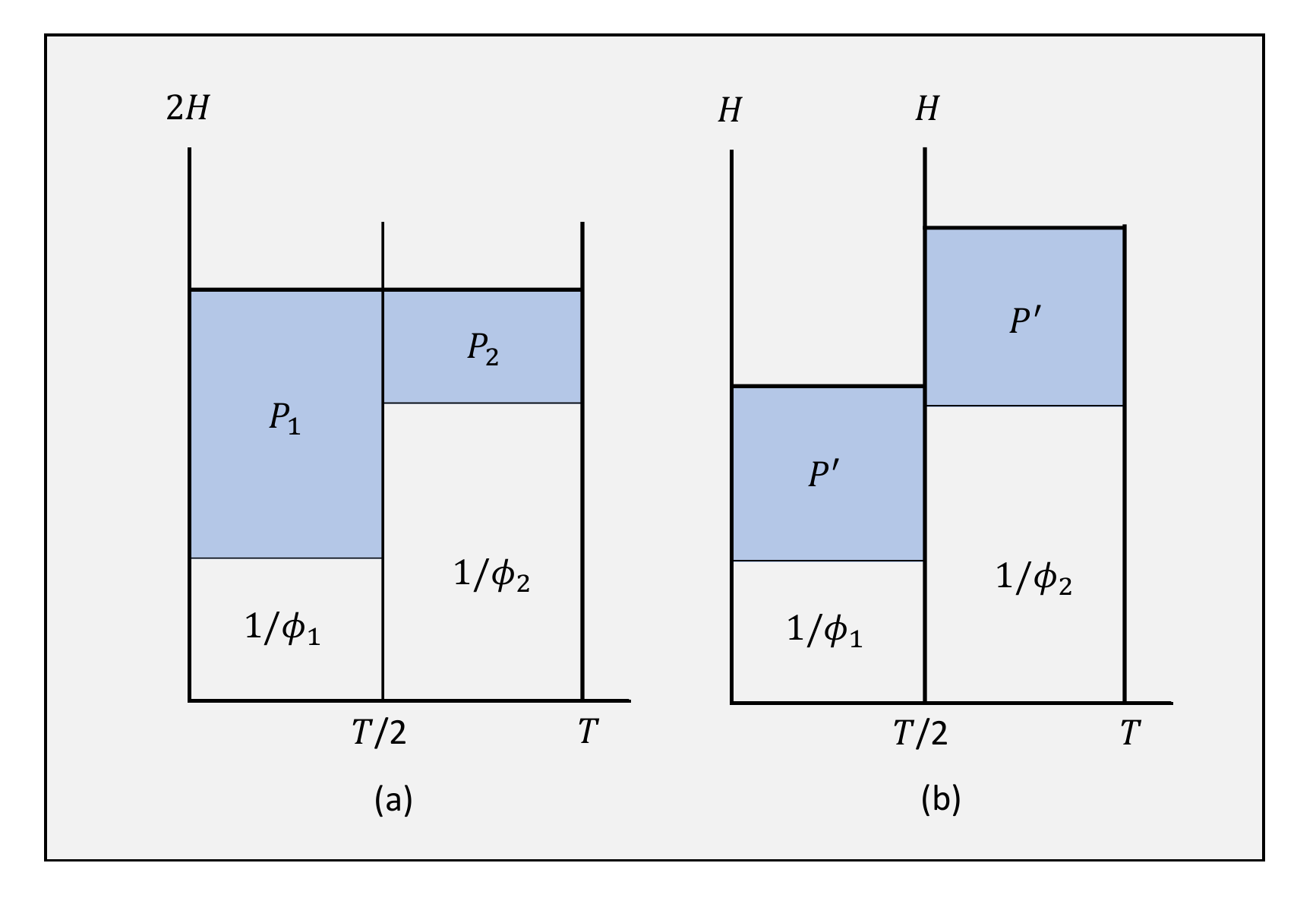}
\caption{Power allocation scheme over time-varying channel with $\phi_1 > \phi_2$. (a) $2H$
units of energy are harvested at instant 0; (b) $H$ units of energy are harvested at instants 0 and $T/2$ \cite{Gunduz14design}.}
\label{direct_water_filling}
\end{figure}

Rules 1 and 2 still hold after replacing ``transmit power''  with ``water-level'' \cite{Tutuncuoglu2012Optimum}. Here the water-levels are defined as $w_t:=P_t+1/\phi_t, t=1,2,\cdots$, where $\phi_t$ are the channel power gains of slotted time-varying channels.
Consider first a simple case consisting of two slots with duration $T$, where $\phi_1 > \phi_2$. The transmit powers can be optimally arranged using a well-known water-filling algorithm, as shown in Fig. \ref{direct_water_filling}(a). The blue regions denote the energy allocated to the transmitters during each slot, and the height of a region, $P_t$, is the transmit power for that slot. The celebrated water-filling algorithm always assigns higher transmit powers to the channels with stronger channel gains.
For a general EH process, applying the water-filling algorithm is not straightforward.
Assume that energy of amounts $H$ is harvested at time instants 0 and $T/2$, respectively. Different from the case in Fig. \ref{direct_water_filling}(a), no more than $H$ units of energy can be allocated to the first slot due to the energy causality. The optimal allocation under energy causality is called directional water-filling \cite{Tutuncuoglu2012Optimum}; see Fig. \ref{direct_water_filling}(b).
The approach is generalized to the broadcast channels in \cite{Ozel2013Optimal}.

\subsubsection{Non-Idle Circuit Power}
In \cite{Ozel2011Transmission} and \cite{Tutuncuoglu2012Optimum}, the circuit power of e.g., converters, filters and mixers, is assumed to be negligible. However, in short-range communications, the circuit power is non-negligible and must be captured in the analysis \cite{Bai2011Throughput,Orhan2012Throughput,Xu2012Throughput}.
When the circuit power consumption is negligible, the transmitter keeps active. Nevertheless, when the consumed circuit power is comparable to the transmit power, the optimal transmission policy must include the ``sleep'' periods, during which the transmitter is off. When a non-negligible circuit power is considered,
the total power consumed at the transmitter, denoted by $P_{\text{total}}$, can be written as
\begin{equation}\label{eq1}
P_{\text {total}}=\left\{
\begin{array}{ll}
    \frac{P(r)}{\vartheta_c}+\rho,		&\text{if}~~P(r)>0,\\
    \beta,			&\text{if}~~P(r)=0.\\
\end{array}
\right.
\end{equation}
Here, $P(r)$ is the transmit power which is in general a convex function of transmit rate $r$;
$\rho$ and $\beta$ denote the circuit power consumption when the transmitter is on and off, respectively; and $\vartheta_c$ is the transmit efficiency of the transmitter. In most existing works, it is generally assumed that $\rho>0$ and $\beta=0$, since typically $\rho \gg \beta$; and $\vartheta_c=1$, as $\vartheta_c$ is a scaling factor \cite{Bai2011Throughput,Orhan2012Throughput,Xu2012Throughput}.

Considering the above non-ideal circuit power model, a so-called optimal ``on-off'' transmission policy is achieved in \cite{Bai2011Throughput} by adjusting the ideal-case optimal transmission scheme in \cite{Tutuncuoglu2012Optimum}. A minimum power level $p^*$ can be derived under the non-negligible circuit power $\rho$. If the optimal transmit power in \cite{Tutuncuoglu2012Optimum} is lower than $p^*$, then the transmitter sends data with the power $p^*$, and turns off once the harvested energy is used up \cite{Bai2011Throughput}.

In \cite{Orhan2012Throughput}, the optimal power allocation over time-varying channels is shown to be the solution to a convex optimization problem, and is portrayed as a directional glue-pouring algorithm integrating the glue-pouring recently developed in \cite{Youssef2008Bursty} and the aforementioned directional water-filling algorithm \cite{Ozel2011Transmission}. The minimum amount of power is allocated to each epoch, depending on the channel state and the non-negligible power consumption of the transmitter circuitry. The amount of harvested energy decides whether energy is ``poured'' into part of an epoch at the minimum power level, or allocated to the entire epoch with a higher power level.

In \cite{Xu2012Throughput}, the tradeoff between EE and spectrum efficiency (SE) is considered under non-ideal circuit power. An EE-maximizing power level $P_{ee}$ is obtained by maximizing the amount of data that can be transmitted with a unit of energy.
Specifically, $P_{ee}$ is defined as
\begin{equation}\label{eqee}
\displaystyle P_{ee}:=\arg \max_{P \geq 0}{\frac{R(P)}{P+\rho}},
\end{equation}
where $P_{ee}$ can be efficiently solved through bisectional search, as $\frac{R(P)}{P+\rho}$ is quasi-concave.
Xu and Zhang \cite{Xu2012Throughput} show an optimal solution with a two-phase scheme. The first stage of the approach is an ``on-off'' transmission with the EE-maximizing power allocated to all on-periods, and the second stage is to continuously transmit with a non-decreasing SE-maximizing power. The optimal offline algorithm is later extended to the scenario of multiple additive white Gaussian noise (AWGN) channels, where the above original power allocation problem with multi-dimensional vectors is solved by equivalently solving a problem with only a single one-dimensional scalar optimization variable through nested optimization techniques~\cite{Xu2012Throughput}.

By taking the battery storage loss into consideration, a special pattern of the optimal power is uncovered by combining the Lagrange multiplier method and DP \cite{yuan15}. In specific, as the history of the battery status has a non-negligible influence on the current status, a DP based technique (or method) is proposed to locate the slot for zero battery level in a backward induction manner with an affordable complexity.

\subsection{Moderate Data Arrival}

\subsubsection{Non-Stationary data and energy arrival processes}
Section II-A assumes there are always data available in the buffer for transmission; in a more general scenario, data and energy arrivals can be bursty over time.
The design goal of the optimal schedule can be to minimize the total  transmission time \cite{Jing2012Optimal} or the total consumed energy \cite{Chen2016Optimal,Chen2014ICC,chen2014Globecom,chen2015provisioning}.
Let curves $\tilde{A}_t$ and $\tilde{D}_t$ be the total amount of data arrived and delivered by time $t$, respectively. Based on data causality, curve $\tilde{D}_t$ is always below $\tilde{A}_t$.
The optimal transmission scheme obtained for a heavy data arrival scenario is no longer feasible
as there might not be enough data in the buffer. Moreover, packets can have different deadlines \cite{Chen2016Optimal,Chen2014ICC,chen2014Globecom,chen2015provisioning}.
Let curve $\tilde{D}^{\min}_t$ be the total amount of data that must be transmitted by time $t$.
Then $\tilde{D}_t$ should always lie above $\tilde{D}^{\min}_t$ to guarantee the deadline requirement (or QoS).
The optimal transmission policy must jointly take into account the constraints in the data domain and those in the energy domain.

To better illustrate the impact of intermittent data arrivals on EH-based data scheduling, we consider first the scenario where energy is available at the beginning of the transmission~\cite{Zafer2005A,Wang2013Energy,Nan2014Energy}.
Then, a transmission strategy can be found by equivalently specifying a non-decreasing and continuous function $\tilde{D}_t$ over time.
In \cite{Zafer2005A}, a calculus method is developed to determine the optimal $\tilde{D}_t$ to minimize the transmission energy consumption for delay-sensitive packets over time-invariant channels. It is shown that the value of $\tilde{D}_t$ can be readily optimized by simply tautening a string between the data arrival curve $\tilde{A}_t$ and deadline curve $\tilde{D}^{\min}_t$.
It is concluded in \cite{Zafer2005A} that the optimal transmission strategy in a moderate data arrival scenario (with unlimited energy) obeys the following two rules:
\\

\textit{{\bf Rule 3:} The transmit rate shall only vary when the data buffer is empty (i.e., there are no undelivered data), or when all the deadline-approaching data is transmitted \cite{Zafer2005A}.}

\textit{{\bf Rule 4:} The transmit rate increases only at the instants when the data buffer is empty, and decreases only at the instants when all the deadline-approaching data is transmitted~\cite{Zafer2005A}.}
\\

\begin{figure}
\centering
\includegraphics[width=0.44\textwidth]{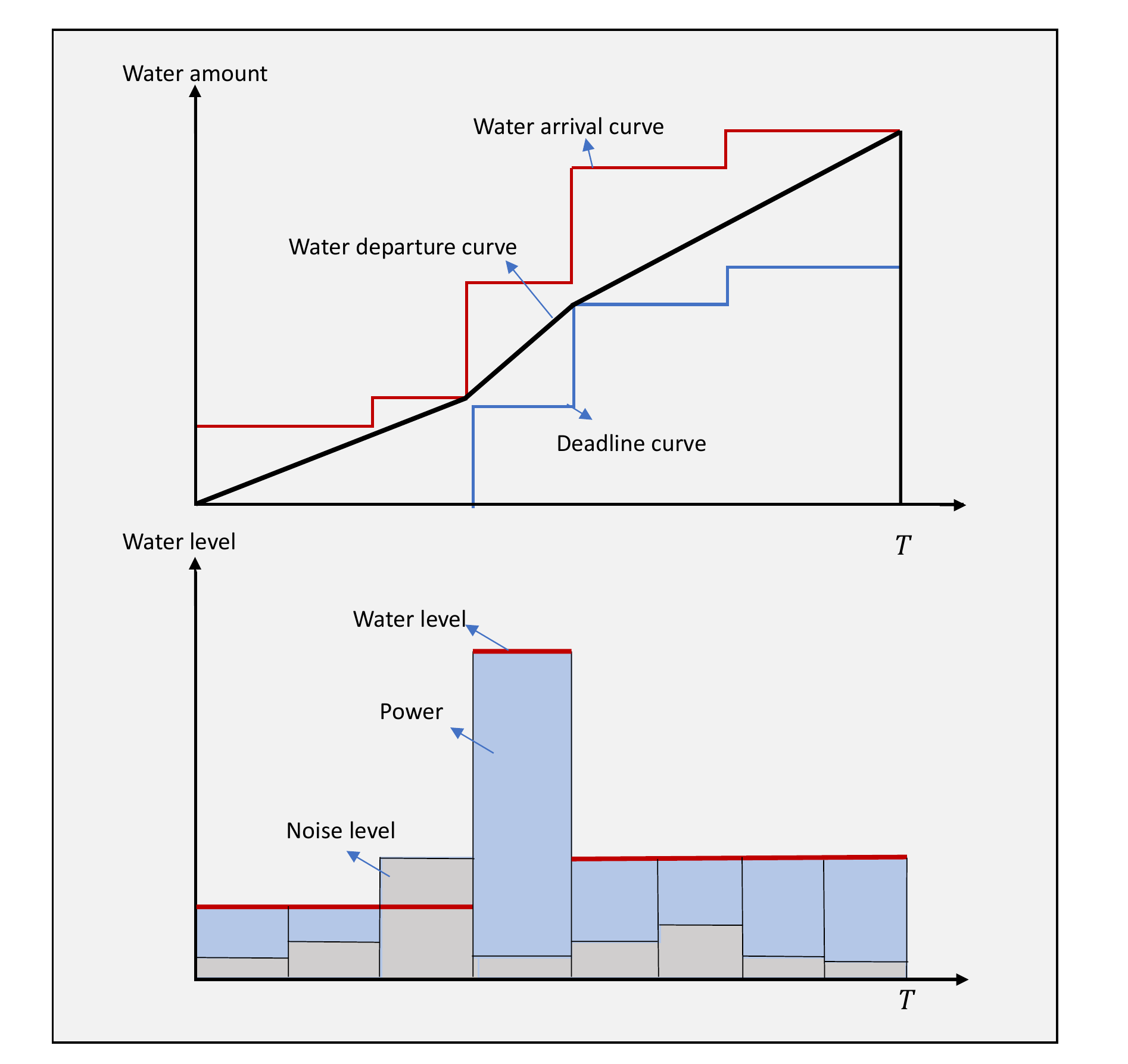}
\caption{(Top) Water arrival curve, deadline curve and optimal water departure
strategy; (Bottom) Water-filling algorithm admitting multiple water levels \cite{Wang2013Energy}.}
\label{water_filling}
\end{figure}

Rules 3 and 4 imply that the visualization method used to construct the optimal transmission policy in \cite{Zafer2005A} has similar procedures as in \cite{Bai2011Throughput,Tutuncuoglu2012Optimum}, which follow Rules 1 and 2. The generalization of Rules 1 and 2 to\cite{Zafer2005A} can be achieved by simply substituting the energy related curves $\tilde{E}_t$, $\tilde{P}_t$ and $\tilde{E}_t-E_{\max}$ in \cite{Bai2011Throughput,Tutuncuoglu2012Optimum} with the corresponding data related curves $\tilde{A}_t$, $\tilde{D}_t$ and $\tilde{D}^{\min}_t$.
By incorporating the calculus method~\cite{Zafer2005A} into the classic water-filling technique, reference \cite{Wang2013Energy} generalizes the optimal rate schedule to time-varying wireless channels. By formulating a convex problem and applying the KKT optimality conditions, an interesting multi-level water-filling pattern is revealed in the optimal schedule with the minimum power consumption in~\cite{Wang2013Energy}, which can be visualized to be the shortest path between the ``water'' arrival and the corresponding minimum ``water'' departure curves, corresponding to the data arrival and deadline curves, respectively, as shown in Fig.~\ref{water_filling}.

Taking non-ideal circuit power into account, the study in~\cite{Nan2014Energy} generalizes the approach developed in \cite{Wang2013Energy}, and proposes an energy-efficient ``clipped string-tautening'' algorithm. It is revealed in~\cite{Nan2014Energy} that the transmitter can take the following three schemes under the optimal data schedule: i) remaining off over the entire slot; ii) transmitting at the energy-efficiency (EE)-maximizing rate for part of the slot; or iii) transmitting at a rate larger than the EE-maximizing rate over the entire slot.

We proceed to consider that both energy and data arrivals are bursty over time.
A so-called ``dynamic string tautening'' technique is put forth to create most energy-savior offline transmit policies even in the presence of a non-ideal transmitter circuit with a considerable power consumption and a limitation of a finite battery capacity \cite{chen2014Globecom}, or without the limitation \cite{Chen2016Optimal}.
The transmit rate $r_t$ and the ``on'' duration $l_t$ per epoch $t$ are the variables to be optimized. Neither of $r_t l_t$ and $P(r_t) l_t$ is convex or concave over $(r_t, l_t)$ in the problem of interest in \cite{Chen2016Optimal}. Yet, the problem can be transformed into a convex one by applying variable substitution with $\Phi_t:=r_t l_t$. For any convex $P(r_t)$, $P(\frac{\Phi_t}{l_t})l_t$ is known to be its perspective, and is convex in $(\Phi_t, l_t)$.

By leveraging convex formulation and its optimality conditions, Rules 1--4 can be met simultaneously, and guide the generation of the optimal strategy by recursively tautening a data departure string in a feasible solution region. The complexity of tautening a string in such a way is low. It is revealed in \cite{Chen2016Optimal} that the current string tautening behavior depends on the past one, as the past data schedule affects the available energy left in the battery.
Instead of tautening two strings (which are the transmitted energy curve following Rules 1 and 2, and the corresponding transmitted data curve following Rules 3 and 4), only the transmitted data curve needs to be optimized since the energy related constraints curves can be translated to the data domain \cite{Chen2016Optimal}.

The algorithm developed in \cite{Chen2016Optimal} is generalized to the online scenario, where the transmission policy is generated in real time. It is also extended to time-varying channels in \cite{Chen2014ICC}. In~\cite{chen2015provisioning}, appropriate models of EH transmitters, with meticulous considerations on the EH rate, deadline requirements, and battery size, are investigated to balance the QoS guarantee and the energy consumption based on ``dynamic string tautening'' algorithms.
Targeting at minimizing the total completion time of transmitting all arrived data, the optimal data schedule is obtained by a recursive visualized scheme which jointly checks the conditions of the data and energy departure curves \cite{Jing2012Optimal} in accordance with Rules 1-4.

Considering joint EH and constant grid power supply, different resource allocation problems are formulated to minimize the total power consumption~\cite{gong13,kang14}, or the grid power consumption~\cite{ahmed13,cui15}.
Through convex formulation and KKT optimality conditions, a two-stage water-filling policy is proposed in a heavily loaded traffic condition with the harvested energy distributed in the first stage and the grid energy supply distributed in the second stage \cite{gong13}. For a moderate traffic arrival case, a multi-stage water-filling strategy is developed to reversely allocate the energy from the last frame to the first frame. This policy can only be generated offline. Ahmed et al.~\cite{ahmed13} develop a stochastic DP algorithm for online power allocation. To bypass the high complexity of DP, a suboptimal online scheme is developed through convex formulation for a good performance-complexity tradeoff~\cite{ahmed13}.

\subsubsection{Stationary data and energy arrival processes}
The celebrated Lyapunov optimization framework has been applied to obtain online dynamic resource allocation with i.i.d. data and energy arrivals, under data queue stability constraint \cite{cui15}.
The Lyapunov optimization technique provides a very effective tool to stabilize and optimize networks of queues. Let $L(t)$ denote the Lyapunov function, which is a non-negative metric to measure queue lengths. The growth of $L(t)$ indicates that the queueing system becomes increasingly unstable.
Consider a system of $I$ queues with lengths of $(Q_1^t, Q_2^t, \ldots, Q_I^t)$. The arrival process of each queue is assumed to be stationary (stochastic). The Lyapunov function can be formulated as $L(t)=\frac{1}{2} \sum_{i=1}^I (Q_i^t)^2$.
Then, $\triangle L(t)= L(t+1)-L(t)$ defines a Lyapunov drift, which is the difference of the Lyapunov functions at two adjacent slots. A feasible way to preserve the system stability is to minimize the Lyapunov drift at every slot and prevent the queue lengths from growing~\cite{Nee10}.

The term $\triangle L(t)+V p(t)$ is defined as the Lyapunov drift-plus-penalty, where $p(t)$ stands for a penalty function, and $V$ specifies a positive weight.
By minimizing the upper bound of $\triangle L(t)+V p(t)$ per slot, one can stochastically minimize the time-average penalty with asymptotic optimality while preserving the system stability.
In \cite{cui15}, the data queue in the transmit buffer and the energy queue in the battery are mapped to $\boldsymbol{Q_i^t}=\{Q_{d,i}^t,Q_{e,i}^t\}$, while the total power consumption specifies the penalty function $p(t)$.
As a consequence, minimizing the maximum value of this Lyapunov drift-plus-penalty per slot leads to the asymptotic minimality of the time-average energy cost, while stabilizing all the data and energy buffers.
By exploiting a continuous-time approximation, DP, and sample-path approach, an analytic framework is introduced to study the power-delay tradeoff relationship in the small delay regime \cite{cui15}.

In \cite{blas}, the EH communication system is assumed to be a finite Markov decision process (FMDP) \cite{bellman1957markovian}, where the energy and data arrivals are Markov processes.
A reinforcement learning-based approach, Q-learning, is developed for the transmitter operation of the EH communication system. For any given FMDP, Q-learning can select the optimal actions and maximize the expected total reward for each and all consecutive steps. 
In \cite{blas}, the transmitter is designed to learn the optimal transmission strategy gradually by taking exploratory actions (i.e. dropping or transmitting the incoming packets) and maximizing the expected sum rewards (i.e. the total throughput). It is shown that the proposed approach asymptotically converges to the global optimum, as the learning time increases.

\section{Energy Harvesting-Based Multipoint-to-Point Wireless System}\label{sec.multiaccess}

\begin{figure}
\centering
\includegraphics[width=0.47\textwidth]{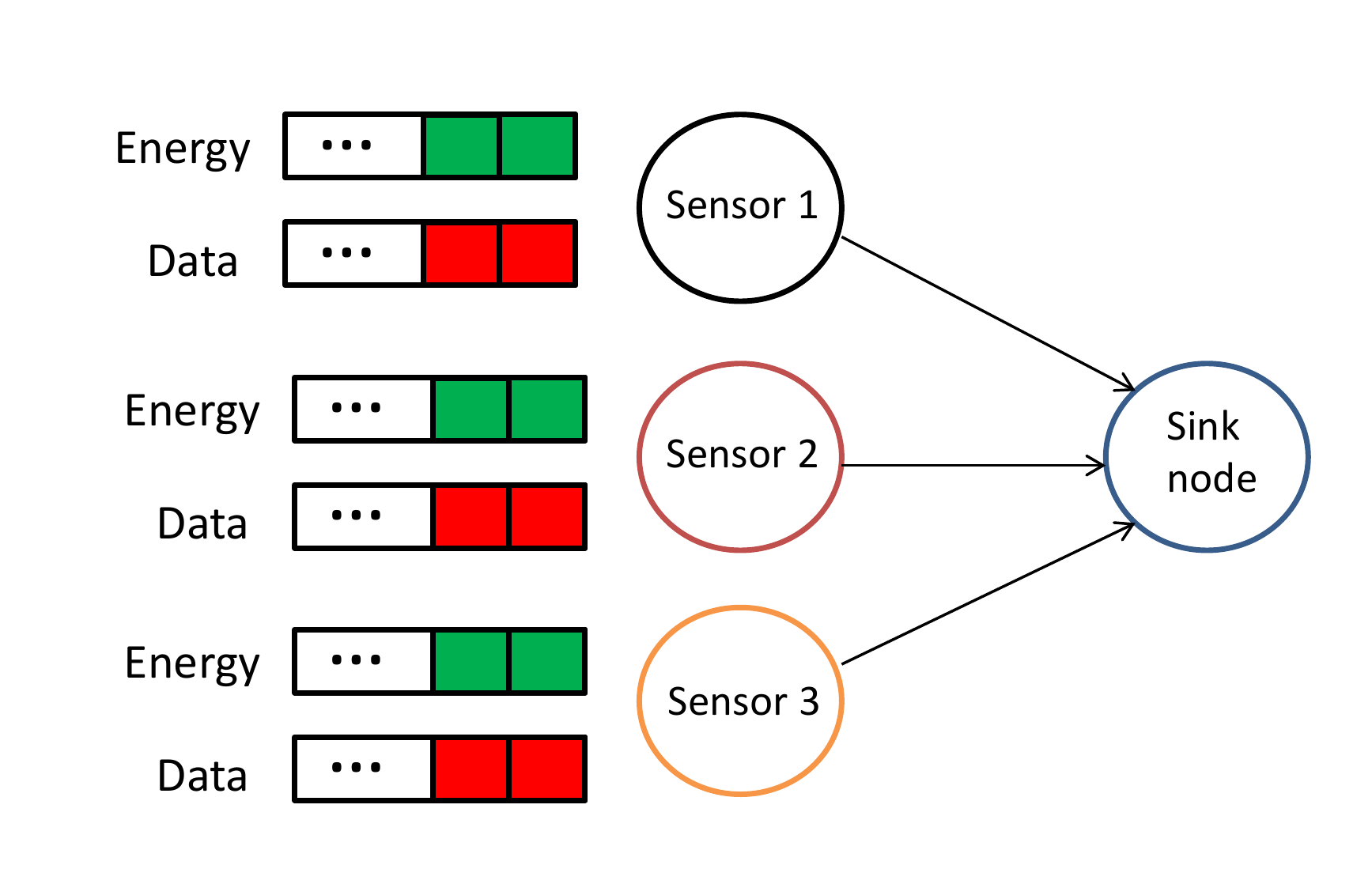}
\caption{Diagram of an EH-based multi-point-to-point wireless communication system.}
\label{multi access}
\end{figure}

Another wireless communication architecture extensively studied in the context of RES/EH is multipoint-to-point network architecture, where there are multiple EH devices (or sensors) and a single sink node; see Fig.~\ref{multi access}. The sensors need to be scheduled to send their data to the sink. Energy/EH-aware scheduling is a key differentiating factor of the multipoint-to-point networks
to the point-to-point networks (discussed in Section III)~\cite{yang12optimal}. In the case of centralized scheduling, the sink collects information, such as queue backlog and energy availability, and selects accordingly the devices to transmit~\cite{yang12optimal,Wang2014Iterative}. In the case of distributed scheduling, random access protocols are developed to trigger the transmission of a device based on its own queue and battery states~\cite{Baknina18,miche,zheng2016game,Iannello2012Medium,Gurakan2018Energy,Jeon15}.

\subsection{Centralized Scheduling}

Yang and Ulukus \cite{yang12optimal} extend their previous works on power allocation for EH point-to-point links \cite{Ozel2011Transmission,Jing2012Optimal} or EH broadcast channels \cite{Ozel2013Optimal} to a multiple access communication system. By capturing the interference between the two users and following the aforementioned Rules 1 and 2, the authors of  \cite{yang12optimal} generalize the well-known backward water-filling algorithm to specify the maximum data departure region with a given deadline. An offline transmission strategy is then obtained by decomposing the transmission completion time minimization problem into several convex subproblems, according to the region sequence at energy arrival instants.

By taking the maximum per-slot energy consumption into account, an iterative dynamic water-filling approach is designed in \cite{Wang2014Iterative} to provide the optimal energy schedule for EH multi-access channels and maximize the throughput of the channels. It is shown that, in practice, the convergence can be reached within only a few iterations.

The optimal resource allocation is specifically designed in multi-input multi-output (MIMO) systems powered by smart grid in \cite{hu16con, hu16}
by maximizing the weighted or expected sum-rate.
Relying on an uplink-downlink duality derived information-theoretically \cite{jindal2004},
the downlink MIMO broadcast channel capacity region in the downlink can be equivalently calculated as the union of the uplink multi-access channels capacity regions when the uplink and downlink have to meet the same sum-power constraint.
To derive the optimal power allocation strategy in an offline fashion,
the Lagrangian dual based subgradient method is leveraged in~\cite{hu16con, hu16}
by applying a nested optimization process.
The downlink covariance matrices can be derived from their uplink counterparts using the uplink-downlink duality.

\subsection{Distributed Scheduling}
Baknina and Ulukus \cite{Baknina18} extend the results of \cite{yang12optimal} to the online scenario where the arrivals or availability of ambient energy are typically unpredictable. A distributed fractional power (DFP) policy is proposed and proven to be near-optimal. By assuming that the energy collected by the users is synchronized Bernoulli process, it is shown that the optimally allocated power decreases until reaching the end of the renewal time, and has a pentagon time-average throughput region. It is also shown in \cite{Baknina18} that the correlation of the energy sources is detrimental to the achievable throughput, as the throughput is much larger under asynchronous Bernoulli energy arrivals than it is under synchronized ones.

Michelusi and Zorzi \cite{miche} investigate a multi-access policy maximizing the utility of the EH wireless sensor networks (WSNs), where different EH sensors transmit packets to a fusion center by randomly accessing a shared wireless channel. Each packet has a random utility value. A distributed random access protocol is designed based on a game theoretic framework, where all sensors perform the same strategy.
The work in \cite{miche} is extended to a multi-channel case with delay-sensitive data transmissions \cite{zheng2016game}.
Considering a similar scenario as the one in \cite{miche}, Iannello et al. \cite{Iannello2012Medium} design the medium access control (MAC) by balancing the tradeoff between time efficiency and transmission probability.

A recent work \cite{Gurakan2018Energy} considers a EH-based two-user cooperative multiple access channel, where the two users perform data cooperation by cooperatively establishing and sending common messages, and perform energy cooperation by wireless energy transfer. By leveraging the Lagrange dual based method, the offline energy allocation and transfer policy maximizing the departure region are jointly optimized.
The exact stability region is characterized in \cite{Jeon15} where two EH nodes randomly access the same receiver. Relying on Loynes' theorem \cite{loynes1962stability}, the analysis of the stability region takes into account the effect of limited energy availability, finite battery capacity, and data reception capability. Loynes' theorem indicates that, if the inbound rate is smaller on average than the outbound counterpart, and the incoming and outgoing processes are ergodic, then the queueing system is stable \cite{loynes1962stability}.

To solve the problem in \cite{hu16con, hu16} over an infinite scheduling period by an online algorithm,
Wang et al. \cite{wang2016} rely on the stochastic subgradient method to obtain resource schedules ``on-the-fly''
by suppressing (decoupling) the time-coupling between the variables and constraints.
The random variables are supposed to be i.i.d..
Using stochastically estimated Lagrange multipliers,
this method updates the subgradients with their online approximations based on
the instantaneous decision variables per time slot.
It is proven in \cite{wang2016} that asymptotically optimal and feasible solutions can be achieved in no need of any prior knowledge of underlying randomnesses.
The optimality gap diminishes when the iteration stepsize approaches zero.

When achieving the optimality, the BS purchases more electricity to transmit its data and charge its battery from the smart energy grid if the grid offers a low energy price, and uses the energy stored in battery if the grid asks for a high price.
The BS can even sell some of its surplus energy back into the grid, hence offsetting part of its energy bill.
Compared with a traditional grid-powered BS, the proposed system in \cite{wang2016} can achieve higher sum rates under the same cost budget.

\begin{figure*}[th]
\centering
\includegraphics[width=0.7\textwidth]{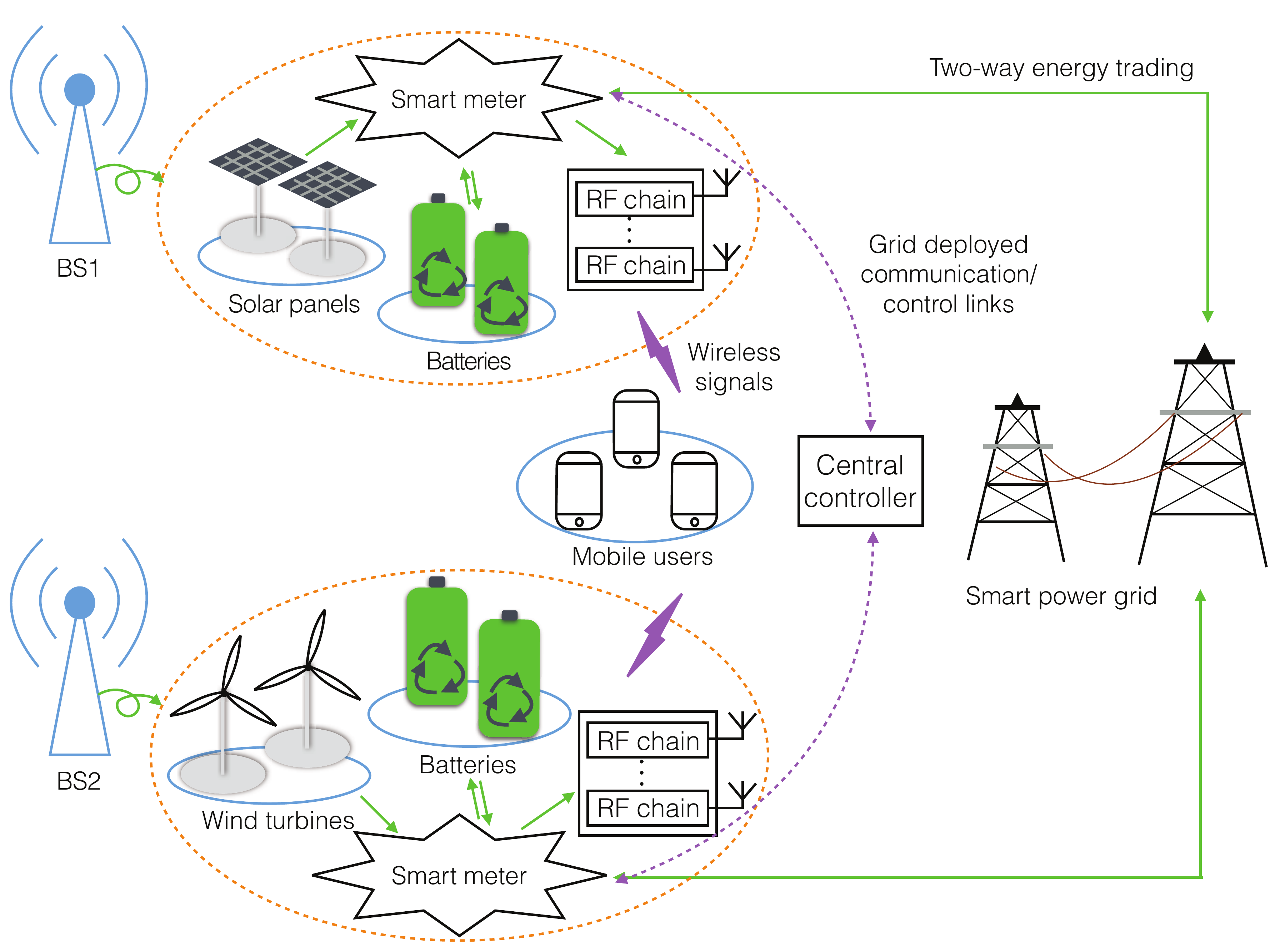}
\caption{A typical smart grid-powered CoMP system with two BSs.}
\label{compmodel}
\end{figure*}

\section{Energy Harvesting-Based Multipoint-to-Multipoint Wireless Communication System}\label{sec.comp}

Coordinated multi-point transmission and reception (CoMP) was initiated in
the Third Generation Partnership Project (3GPP), as one of the key technologies
in the long-term evolution-advanced (LTE-A) standard,
in order to achieve interference management and mitigation.
A CoMP system can be
{\red viewed as a group}
of geographically collocated transmit antennas
which coordinately serve several multi-antenna end users \cite{lee12}.
Thus, a CoMP system can be referred to as a multipoint-to-multipoint system.
For coordinated transmission in the downlink,
{\red the data sent by several transmission points are coordinated to enhance the
acquired intensity of the expected information
at the user equipment (UE) or to decrease the inter-cell interference.}
For coordinated reception in the uplink, it is ensured that the uplink
{\red data from the UE can be steadily detected}
by the network with limited uplink interferences and the existence of multiple reception points.

As there are a large number of users and service demands, it costs great energy to realize transmit beamforming, interference alignment, user scheduling,
and backhaul signaling in multipoint-to-multipoint systems.
The integration of RES can reduce the electricity consumption from traditional power sources and
facilitate economic and ecological operations of such systems in the age of ``green communications''.

Many recent works have minimized the energy consumption and operational cost of RES-powered
multipoint-to-multipoint systems.
Classic techniques and methodologies applied in such problems include the Lagrangian dual-based method,
(stochastic) subgradient method (SGM) \cite{sgm86, sgm09, sgm17, Kiwi04}, cutting-plane method (CPM) \cite{cpm95, cpm02, cpm15},
and proximal bundle method (PBM) \cite{Kiwi95, Kiwi, pbm06}.
For a minimization problem, its Lagrangian dual function is always concave,
{\red even if the primal problem is not convex \cite{boyd}.}
The optimal dual solutions can be readily obtained by using off-the-shelf convex optimization tools,
such as SDP, SOCP, and interior point method \cite{boyd}.
To effectively solve non-differentiable convex dual problems, SGM and CPM are commonly used,
as they deal with subgradients, rather than gradients, of the objective functions.
Although CPM converges faster (i.e. with fewer iterations) than SGM, it is computationally more demanding (per iteration),
and does not allow for a distributed implementation \cite{Xiao04}.
The stochastic SGM is applied when the random variables are i.i.d. and traditional online solvers,
such as DP, are intractable
{\red since the optimization variables are closely coupled over time.}
The optimal primal solutions can be recovered from the dual solutions with no duality gap when the primal problem is strictly convex.
Extra care must be taken for such operations
{\red if the original problem is not strictly convex.}
A sophisticated method to deal with this situation is the PBM,
{\red which is to approach the epigraph of a function
through using the intersection of several halfspaces.}


\subsection{Minimization of Total Energy Consumption}
A CoMP downlink energized by a smart power grid
is studied in~\cite{Xu16} where the BSs have on-spot RES
and carry out bidirectional energy trading in real-time with the smart grid.
Yet, the BSs may not be equipped with energy storage capability.
The optimal solution for minimizing the
{\red overall energy expenditure is obtained by an approach utilizing convex optimization}
and uplink-downlink duality \cite{yu2007tx}.
In particular, let $k \in {\cal K}=[1, \ldots, K]$ denote the user index, and $\mathbf{w}_k$ the beamforming vector
associated with user $k$.
A phase rotation is performed for each $\mathbf{w}_k$,
since the phase rotation does not change the signal-to-interference-plus-noise ratio (SINR) constraints,
i.e., $\text{SINR}_k(\{\mathbf{w}_k\})=\text{SINR}_k(\{\mathbf{w}_k e^{j\phi_k}\})$,
where $\phi_k$ is an arbitrary phase.
Without loss of optimality, $\{\mathbf{w}_k\}$ is chosen in such a way that
$\mathbf{h}_{k}^H \mathbf{w}_k$ is real, and $\mathbf{h}_{k}^H \mathbf{w}_k \geq 0$ in \cite{Xu16},
to convert the original nonconvex SINR constraints:
\begin{equation}
\frac{|{\mathbf{h}_{k}^H} \mathbf{w}_k|^2}{\sum_{l\neq k} (|{\mathbf{h}_{k}^H} \mathbf{w}_l|^2) + \sigma_k^2} \geq \gamma_k,
~\forall k \in {\cal K}
\end{equation}
into convex (second-order conic) constraints \cite[eq. (14)]{Xu16}:
\begin{equation}
\sqrt{\sum_{l \in {\cal K}} |\mathbf{h}_k^H \mathbf{w}_l |^2+ \sigma_k^2} \leq \sqrt{1+\frac{1}{\gamma_k}} \mathbf{h}_k^H \mathbf{w}_k,
~\forall k \in {\cal K},
\end{equation}
where $\mathbf{h}_k$ stands for the channel vector of any user $k$.

Xu and Zhang \cite{Xu16} exploit the uplink-downlink duality to convert
the multiple-input single-output broadcast channel (MISO-BC)
to a dual single-input multiple-output multiple access channel (SIMO-MAC) by
taking conjugate transpose of the channel vectors,
under the same SINR constraints $\{\gamma_k\}$.
Hence, the optimal transmit beamformers $\{\mathbf{w}_k^*\}$ can be obtained by first deriving the uplink dual problem solution
via an iterative function evaluation procedure \cite{wiesel2006},
and then mapping the solution to the original problem.
Other variables of the problem are decided via an ellipsoid method \cite{boyd2}.

As the optimal solver incurs a high computational complexity,
Xu and Zhang \cite{Xu16} also propose a suboptimal solution of a comparatively lower complexity,
by implementing zero-forcing (ZF) beamforming methods at the BSs.
The transmit beamforming vectors are produced to
cancel any mutual interference between different users,
i.e., $\mathbf{h}_k^H \mathbf{w}_l =0$, where $l,k \in {\cal K}$ and $l \neq k$
are the indexes to two different users.
{\red As typically required, ZF beamforming is only implementable under the condition
that there are fewer users than the transmit antennas across the BSs.}

\subsection{Minimization of  Operational Cost}
{\red Due to an unbalanced supply of RES and energy requirements across geographically dispersed BSs,
and a considerable price gap of the BSs purchasing (or selling) electricity from (or towards) the smart energy grid,
it is cost-effective for these BSs to collectively plan their transactional dealings with the smart energy grid,
as well as electricity usage for CoMP-enabled communications.}
A framework is developed in \cite{wang2015} to capture finite storage, bidirectional electricity trading
and dynamic pricing of the smart grid into CoMP downlink communication systems
with imperfect channel state information (CSI) at the transmitter.
The system model is pictorially illustrated in Fig. \ref{compmodel},
where each BS has RES devices (e.g. solar-electric converters and/or hydroelectric generators),
local energy storage devices (e.g. batteries),
and a smart meter to collect energy information and coordinate bidirectional electricity trading activities
by interacting towards the smart power grid.
The BSs in the CoMP cluster jointly serve the mobile users.
In such a system, a central controller is necessary to collect the information across the entire network,
and correspondingly coordinate the activities of the BSs.

{\red The authors of \cite{wang2015} develop the worst-case energy scheduling and transmit beamforming schemes
to minimize the system-wide transaction expenditure,
while guaranteeing user QoS in a CoMP-enabled downlink network.
Let $P_{b,i}^t$ stand for the power transferred into, or taken out of,
the batteries at slot $t$ for BS $i$.}
In the case of $P_{b,i}^t>0$, the BS charges the battery.
If $P_{b,i}^t<0$, the battery is being discharged.
$P_{g,i}^t$ denotes the total power usage for BS $i$ at time slot $t$ (including the transmit power with respect to
beamforming vectors and constant circuits consumption).
With the auxiliary variable $P_i^t=P_{g,i}^t+P_{b,i}^t$ and the energy transaction prices $\alpha_t$ (buying) and $\beta_t$ (selling),
{\red the worst-case transaction expenditure of BS $i$ across the entire
scheduling period can be formulated as \cite{Zha2012} }
\begin{equation}
G(\{P_{i}^t\}):= \max_{E_i \in {\cal E}_i} \sum_{t=1}^T \Big(\alpha^t[P_{i}^t-E_i^t]^+
- \beta^t [E_i^t-P_{i}^t]^+\Big),
\end{equation}
where $E_i^t$ is the harvested energy of BS $i$ at time $t$,
{\red $[P_{i}^t-E_i^t]^+$ is the shortfall of the energy to be procured from the smart power grid,
$[E_i^t-P_{i}^t]^+$ is the redundant energy to be sold to the grid,}
and $[a]^+:= \max\{a, 0\}$.

Applying the semidefinite relaxation (SDR) technique \cite{luo2010se} and the S-procedure \cite{Polik2007},
the energy management and transmit beamforming
{\red optimization is constructed as a convex problem.}
Due to the non-differentiability of $G(\{P_{i}^t\})$, this nonsmooth convex problem is intractable to general solvers.
Its global optimal solution can be
{\red acquired offline}
by employing the Lagrangian dual based subgradient iteration \cite{Bert99, Bert03, Kiwi04, Xiao04, Gatsis12},
together with a proximal bundle method \cite{Kiwi, Kiwi95}.
Ahead-of-time resource planning can be realized,
{\red provided that the energy and data arrivals are obtained beforehand.}

\subsection{Minimization of  Long-term Average Cost}
The dynamic energy management of a CoMP system is considered in an infinite scheduling horizon in \cite{wang2016dyn}.
{\red The target is to minimize the time-averaged
overall expenditure over an infinite time horizon}
by determining the power allocation variables, i.e.
$\min_{\{P_i^t, P_{g,i}^t, C_i^t\}} ~\lim_{T\rightarrow \infty} \frac{1}{T} \sum_{t=0}^{T-1} \sum_{i=1}^I  G(P_i^t).$
The battery level relations (as one of the constraints)
$C_i^{t+1} = C_i^{t} +P_i^t - P_{g,i}^t, \, \forall i, t$
couple the optimization variables over time.
{\red This makes the original problem hardly malleable for traditional solvers such as DP.}

Assuming that the RES amounts and energy transaction prices $\{\mathbf{e}^t, \alpha^t, \beta^t\}$ are i.i.d.
at every individual slot, the authors of \cite{wang2016dyn} use the Lagrange dual based stochastic subgradient approach to tackle this problem.
In essence, this method updates the Lagrange variables by their stochastic estimates at each time slot,
which can reduce the otherwise high computational complexity.
Based on the stochastic iterations, the authors then construct a virtual energy queue $Q_i^t$ for each BS $i$.
The virtual queue obeys the same dynamic equation as in battery level:
$Q_i^{t+1}=Q_i^t + P_i^t(\mathbf{Q}^t) - P_{g,i}^t(\mathbf{Q}^t)$,
where $P_i^t(\mathbf{Q}^t)$ and $P_{g,i}^t(\mathbf{Q}^t)$ are
{\red procured by tackling the dual problem
with stochastic estimates of the Lagrange variables.
Different from real queues,}
the value of $Q_i^t$ can be negative, which is important to the procedures of the
virtual-queue based online control (VQOC) approach at the central scheduler to obtain the asymptotically optimal solutions.
Depending on the state-of-the-art stochastic optimization methodologies \cite{Urg11, Lak14},
the solution can be proven to be asymptotically optimal over a long term.

{\red A two-scale online resource management problem is
investigated for RES-integrated CoMP-enabled communications in~\cite{wang2016two}.
By considering the dynamics of CSI, RES, beforehand/real-time electricity prices, and battery shortcomings,
a stochastic optimization task is built up to minimize the average electricity transactional expenditure
over a substantially long time frame, and also satisfy the QoS of the users.
The authors of \cite{wang2016two} reformulate the primal problem into a malleable structure}
by replacing the time-coupled queue dynamics with a time-averaged constraint, i.e.
$C_i^{t+1} = \vartheta_b C_i^t+P_{b,i}^t, ~ C_{\min} \leq C_i^t \leq C_{\max}, ~ \forall i$
are replaced by$(1-\vartheta_b)C_{\min} \leq \bar{P}_{b,i} \leq (1-\vartheta_b)C_{\max}, \quad \forall i,$
where $\vartheta_b \in (0,1]$ is the storage efficiency.
It is supposed that the battery capacity $C_{\max}$ is finite and the minimum energy level of the battery is $C_{\min}$.
By doing so, the variables $\{C_i^t\}$ are suppressed in the original problem,
and other optimization variables are ``decoupled''
{\red over time.}
The reformulated problem is a relaxed version of the original problem.

A two-scale online optimization method is then
{\red proposed to create ruling policies in real-time in \cite{wang2016two}}
by minimizing the Lyapunov drift-plus-penalty
$\triangle L_t + V p_t$, where $\triangle L_t = L_{t+1}-L_t$ is the Lyapunov drift,
$p_t$ is a penalty function (in the paper $p_t$ is the time-average energy consumption (or utilization) of the BSs),
{\red and $V$ is a preset positive weighting coefficient of the penalty.
It is analytically validated in \cite{wang2016two} that asymptotically close-to-optimal resource allocation can be established
for the above-mentioned original problem with proper settings,
without any prior statistical knowledge of the underlying stochastic processes.
}

\section{Energy Harvesting-Based Multi-Hop Wireless Communication Link}\label{sec.multihop}

\subsection{Information Cooperation Between Nodes}
\begin{figure}
\centering
\includegraphics[width=0.49\textwidth]{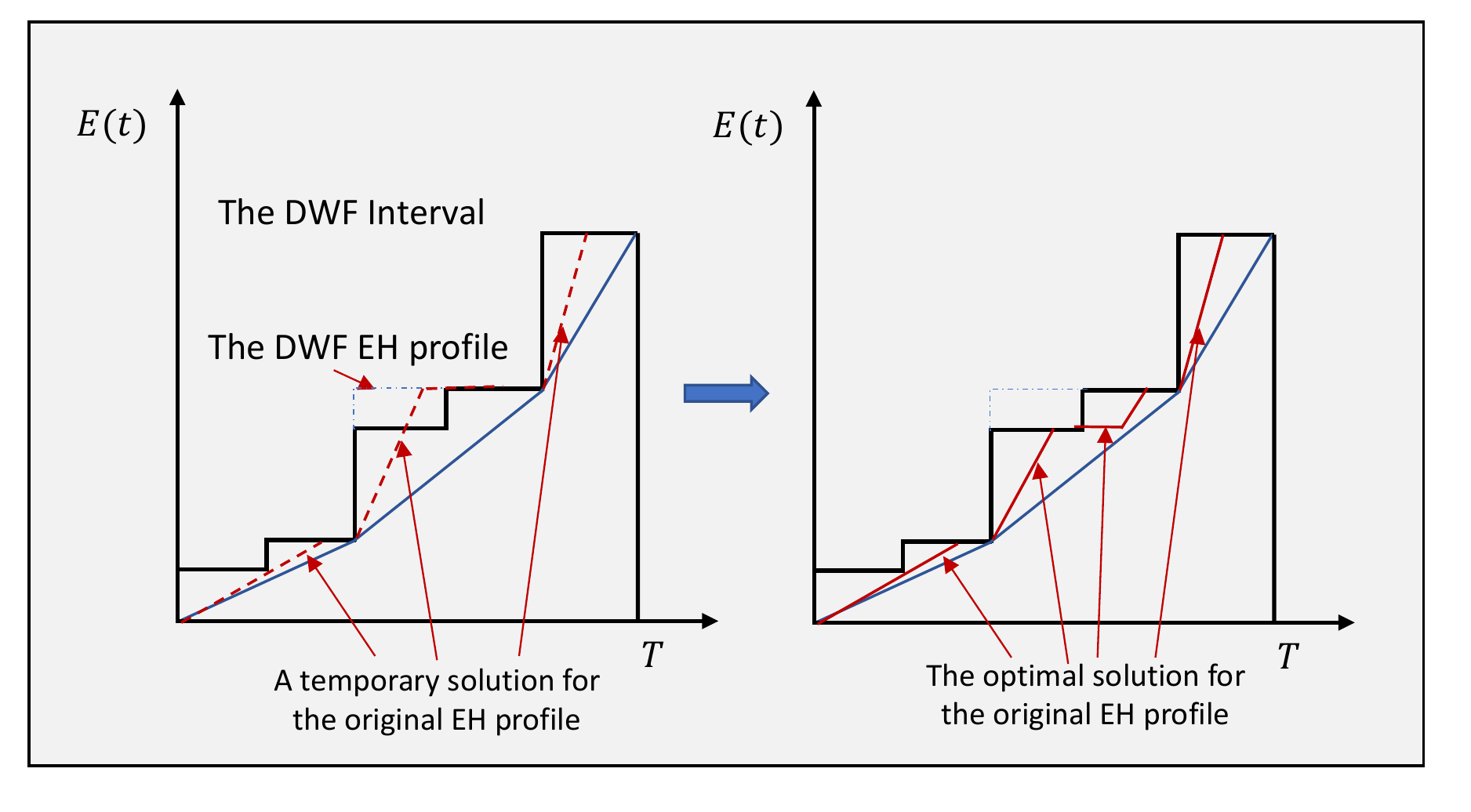}
\caption{The solution to the relaxed EH profile in \cite{luo13} is slightly modified to yield the solution for the original, non-relaxed power allocation problem of \cite{luo13}. The slope of the solution is the optimal transmit power at every instant.}
\label{luo13}
\end{figure}

Cooperation between nodes is an effective way to enlarge system throughput and improve diversity for wireless communication systems. Luo et al. \cite{luo13} consider a dual-hop network with an EH-powered source node and a half-duplex relay node powered by persistent power supply, and investigate joint time and power management.
An important insight is that directional water-filling (DWF) developed for single-hop communication
systems \cite{Ozel2011Transmission} may be generalized to this two-hop scenario \cite{luo13}.
Specifically, based on DWF, a performance upper bound is first found for any given energy arrival process at the source.
The properties of the optimal solutions are then derived, which indicate that: i) The source (relay) node should employ the same transmit power during a given DWF interval; and ii) The data buffer at the relay and the energy buffer at the source are emptied at the end of DWF intervals. The resultant relaxed EH profile
is later modified to optimize the time and power allocation, as shown in Fig.~\ref{luo13}.

The ``on-off'' strategy during each interval is strongly analogy to the problem of sum-rate maximization in the presence of  battery leakage \cite{Devillers2012A} or with non-ideal circuit power consumption \cite{Bai2011Throughput,Orhan2012Throughput,Xu2012Throughput}. However, the ``on-off'' structure in \cite{luo13} results from a half-duplex constraint.
The structure is attributable to the objective of the optimization problem and the constraining factor of battery leakage in \cite{Devillers2012A} and non-negligible circuit energy consumption
in~\cite{Bai2011Throughput},~\cite{Orhan2012Throughput,Xu2012Throughput}.

Huang et al. \cite{huang13} consider EH source and relay nodes performing decode-and-forward (DF) relay, with the a-priori knowledge of the EH profile.
In the case of transmitting delay-constrained (DC) traffic, a new power allocation strategy is developed to maximize the throughput by using the KKT optimality conditions. It is shown in \cite{huang13} that the search algorithm over the two-dimensional EH profiles of the source and relay is an extension of the earlier algorithm developed over the one-dimensional EH profiles in \cite{Jing2012Optimal}.
In the case of transmitting non-delay-constrained (NDC) traffic, based on a separation principle \cite{huang13}, the original problem can be solved by two stages: the transmit power of the source is first jointly optimized by ignoring the requirement of the relay, according to which the transmit power of the relay is then optimized.
It is noted in \cite{huang13} that, in practical EH scenarios, ``energy diversity'' exists due to the independent EH processes at the source and the relay. By relaxing the decoding delay, the proposed transmission policy for NDC traffic can exploit ``energy diversity'' in cooperative communication, and result in a much larger system capacity than the DC counterpart.

Pappas et al. \cite{pappas} study the non-negligible effect of EH on a two-hop network with a collision-prone wireless channel, e.g. IEEE 802.11 WiFi, where an EH-powered source node and an intermediate relay node both receive external traffic. A cooperation of the source and the relay is performed at the protocol (network) level, where the relay takes responsibility of data transmission and can decode the transmissions of the source.
By deriving the sufficient and necessary conditions for the stability of the traffic queues, tight inner and outer bounds of the stability region are obtained with a given transmission probability.

\subsection{Information and Energy Cooperations Between Nodes}
Unlike \cite{luo13,huang13} and \cite{pappas} where energy cooperation between nodes is largely overlooked, Gurak et al. \cite{gurak} develop an iterative algorithm which jointly optimizes power control,
data routing and energy transfer for a EH communication network. In \cite{gurak}, all nodes can harvest energy from ambient environments, and transfer part of the harvested energy to neighboring nodes by energy cooperation.
Based on the Lagrangian function of the convex energy management problem, the necessary conditions are established for the optimal solution. The proposed algorithm is shown to converge towards a Pareto-optimal equilibrium. When there is no energy cooperation, it is shown in \cite{gurak} that, with fixed data flows, a higher transmit power needs to be assigned towards the links either suffering from stronger noises or admitting higher data flows. It is also revealed in \cite{gurak} that, with multiple attempts of EH per node, the optimal power consumption of the outgoing links per slot is equal to that of point-to-point transmission. 

\begin{figure}
\centering
\includegraphics[width=0.47\textwidth]{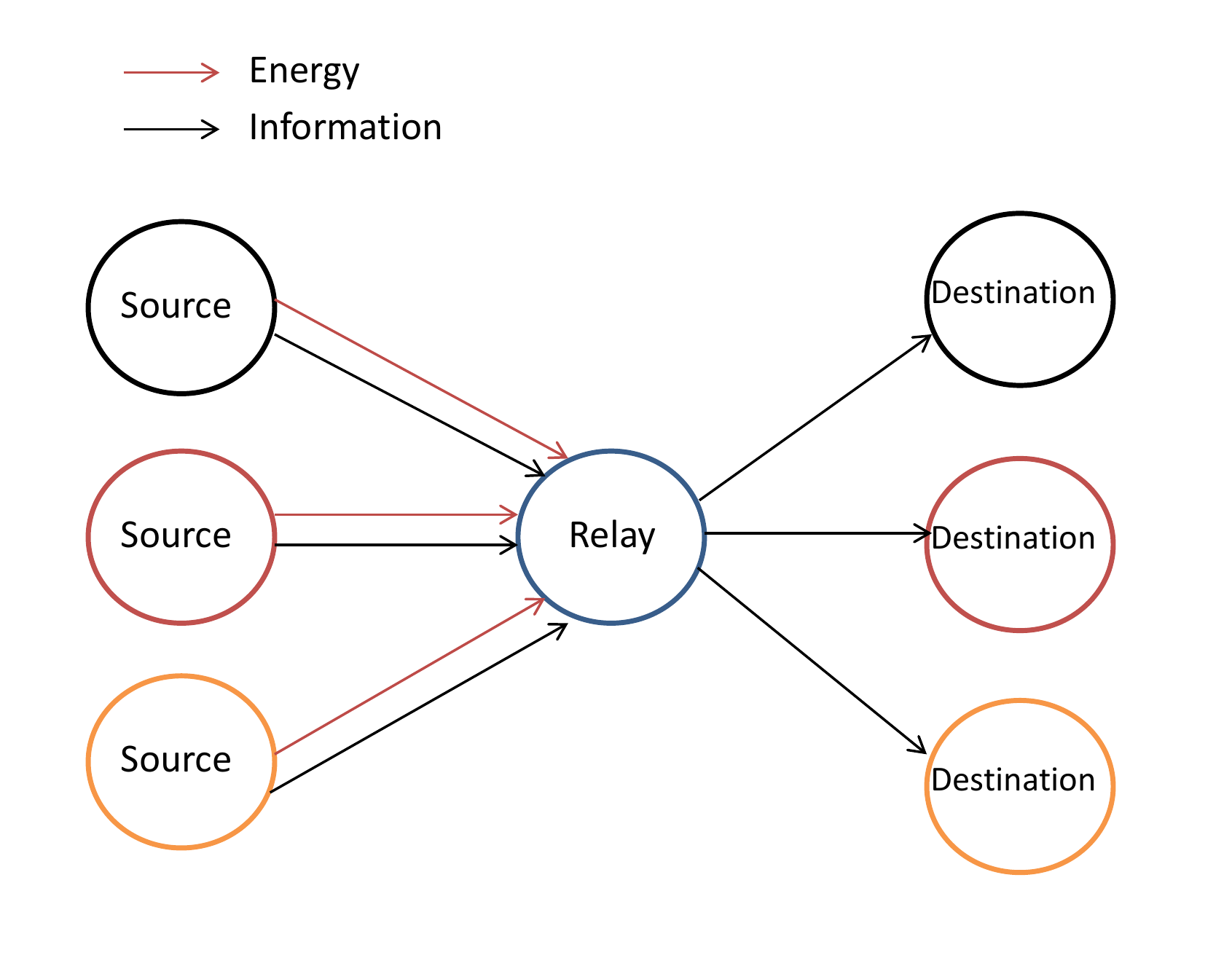}
\caption{Cooperative communication using EH relay.}
\label{ding14}
\end{figure}

In \cite{Ding2014Power}, the power allocation strategies and the resultant outage probabilities are studied in a cooperative communication system, where multiple source-destination pairs are connected by one single EH relay, as shown in Fig.~\ref{ding14}.
By exploiting simultaneous wireless information and power transfer (SWIPT) technologies, the source nodes transmit signals and energize relaying retransmissions.
Assuming that CSI is available at the transmitter, a centralized strategy is proposed based on the concept of sequential water-filling. Specifically, the strategy always serves users with better channel conditions first. Such an opportunistic scheme can minimize the outage probability of the system. A distributed auction based strategy is developed to balance between the overall system performance and the computational complexity, where the relay updates the power allocation scheme at each iteration after the destination nodes submit their bids to the relay.

\section{Wireless Power and Information Transfer}\label{sec.wpt}
So far, energy is harvested from ambient environments (e.g. solar power and wind power)
and treated often as constraints in the optimization of wireless networks.
{\red The approach arising from wireless power transfer (WPT) is able to}
provide a convenient and flexible way for EH that can be
performed anywhere, at anytime, under any weather condition, and for any desirable amount.
{\red Practically, WPT can be realized by multiple different technologies,}
such as induction, magnetic resonating, and electromagnetic (EM) radiation.

As the low-power Internet-of-Things (IoT) devices such as sensors and tags proliferate in the 5G/B5G wireless networks,
how to power end users with green energy has become a critical issue for system designs.
{\red WPT has arisen as a new and promising technique to offer on-spot and on-demand energy replenishment to wireless networks.
Since radio signals can transfer power and information simultaneously,}
study on SWIPT has been pursued for wireless communication systems.

\subsection{SWIPT-Enabled Single-Output  Systems}
Earlier designs on SWIPT systems focus on single-input single-output (SISO) settings.
In particular, a SISO wireless link is considered in \cite{liuliang} where the receiver
{\red is short of persistent power sources and has to restock energy through}
WPT from the signals sent by the transmitter.
{\red The single-antenna receiver cannot decipher information and collect energy}
independently from the same communication signal.
A dynamic power splitting (PS) scheme is developed to
{\red separate the communication signal into two fluids with adaptable powers
for information deciphering and EH, respectively, according to the CSI known at the receiver.}
The PS factors (PSFs) of the receiver and transmit power are
{\red collectively optimized in \cite{liuliang} to achieve the largest ergodic capacity,}
satisfying the demand for EH amount.
Given the special structure of the non-convex optimization problem, the Lagrange duality
{\red approach is employed to procure}
the globally optimal solution.

A Pareto-optimal algorithm is proposed in \cite{tdd} to optimally pick a terminal device and distribute its power budget
across the orthogonal frequency division multiplexing (OFDM) subcarriers under an SISO setting.
The maximum EE is attained in both of the forward and reverse link directions in \cite{tdd},
{\red with known PSFs and uplink/downlink operating time.}
The sum rate in the uplink is maximized in \cite{har-then-tran} by optimizing the durations of both the uplink and the downlink,
where the terminals are attended one after the other in both of the uplink and downlink.
Combined
power allocation and time switching (TS) policy is developed in \cite{tang19} for an SISO NOMA system.
The EE of the network is maximized while
{\red meeting the demands on transmit power budget, data rate, and EH amount per user.
A two-layer approach with the Dinkelbach method~\cite{Ished} is put forth to solve the problem.}

Some existing works are particularly interested in the downlink of MISO SWIPT
{\red networks, and give no consideration to the uplink.
Given an access point (AP) installing multiple transmit antennas and the PSFs of multiple users each
operating a single receive antenna,
transmit beamforming techniques are investigated}
in \cite{6783665} and \cite{jointPS} in attempts to minimize the downlink transmit power of the multi-antenna AP,
while satisfying some given requirements of the transmit rate and EH amount.
{\red SDR is applied to convert the original task into a convex program,
and suboptimal solutions are obtained by leveraging}
ZF \cite{6783665} and SINR maximization \cite{jointPS}.
Interference alignment is adopted in \cite{optimal} to
{\red subdue interference among information transmission
and achieve the totally minimized transmit power in the downlink.}
The semidefinite programming (SDP) technique is applied in \cite{robust1,robust2,robust3} to extend \cite{jointPS} and \cite{optimal}
to enhance robust beamforming under imperfect CSI.
The EE of an SWIPT system is maximized in \cite{A1}
{\red by considering a persistent circuit power at the AP, as well as the terminal stations.
Considerable circuit power is accounted only at the transmit node in \cite{A2},}
without addressing the receiver side.

Harvest-then-transmit policies have been applied to multiuser MISO systems,
{\red where only the power is delivered through the downlink stream (as contrary to the transfer of both power and information in SWIPT).
In this scenario, power splitting at users would not be called for.}
Combined time splitting and beamforming design are considered in \cite{shenchao} to maximize the sum throughput.
{\red In the scenario of perfect CSI at the transmit node, a semi-closed-form solution is developed
by utilizing the strict concavity of the problem.
Taking into account Gaussian CSI errors,
a robust approach is developed to maximize the sum throughput
given a channel outage probability requirement.}
The downlink energy beamformers
{\red and the information transmit power and beamformers in the uplink
are collectively adjusted to achieve the maximum system throughput in \cite{MultiAntenna},
where all users can send data at the same time.}
A spectral radius minimization problem is
{\red constructed and tackled in \cite{MultiAntenna} by leveraging the non-negative matrix theory.
Generalized eigenvectors are applied
to find the optimal beamformers and link operating time in \cite{VT}.}
The combined uplink and downlink sum rate is maximized in \cite{7874074}, without considering QoS in the downlink.

Non-linear EH model is considered in \cite{tran18} and \cite{zhou18} for a multiuser MISO system with SWIPT.
In \cite{tran18}, issues on power-efficient, user-fairness, and channel non-reciprocity are addressed
by a multi-objective optimization problem.
{\red To secure the primary system, an artificial-noise-aided collaborative jamming policy is developed in \cite{zhou18}.}
Transmit power is minimized under secrecy rate and EH constraints.
Algorithms based on SDR or a cost function are developed for the problem.


\subsection{SWIPT-Enabled MIMO System}
\begin{figure}
\centering
\includegraphics[width=0.45\textwidth]{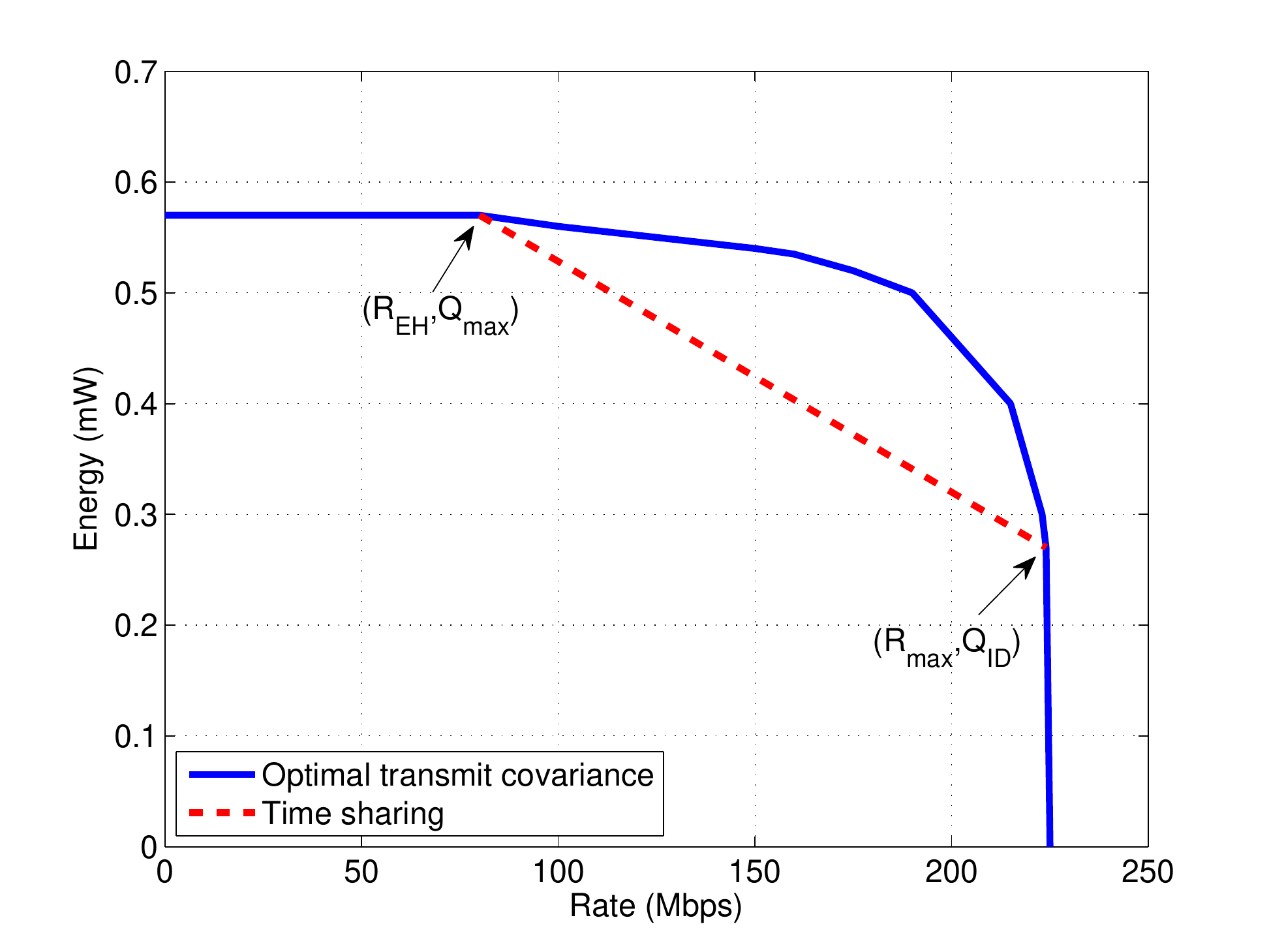}
\caption{The R-E region of a MIMO BC with two separate receiving devices for energy and information, where $M =N_{EH} =N_{ID} =4$
\cite[Fig. 4]{rui13}.}
\label{reseparate}
\end{figure}

\begin{figure}
\centering
\includegraphics[width=0.45\textwidth]{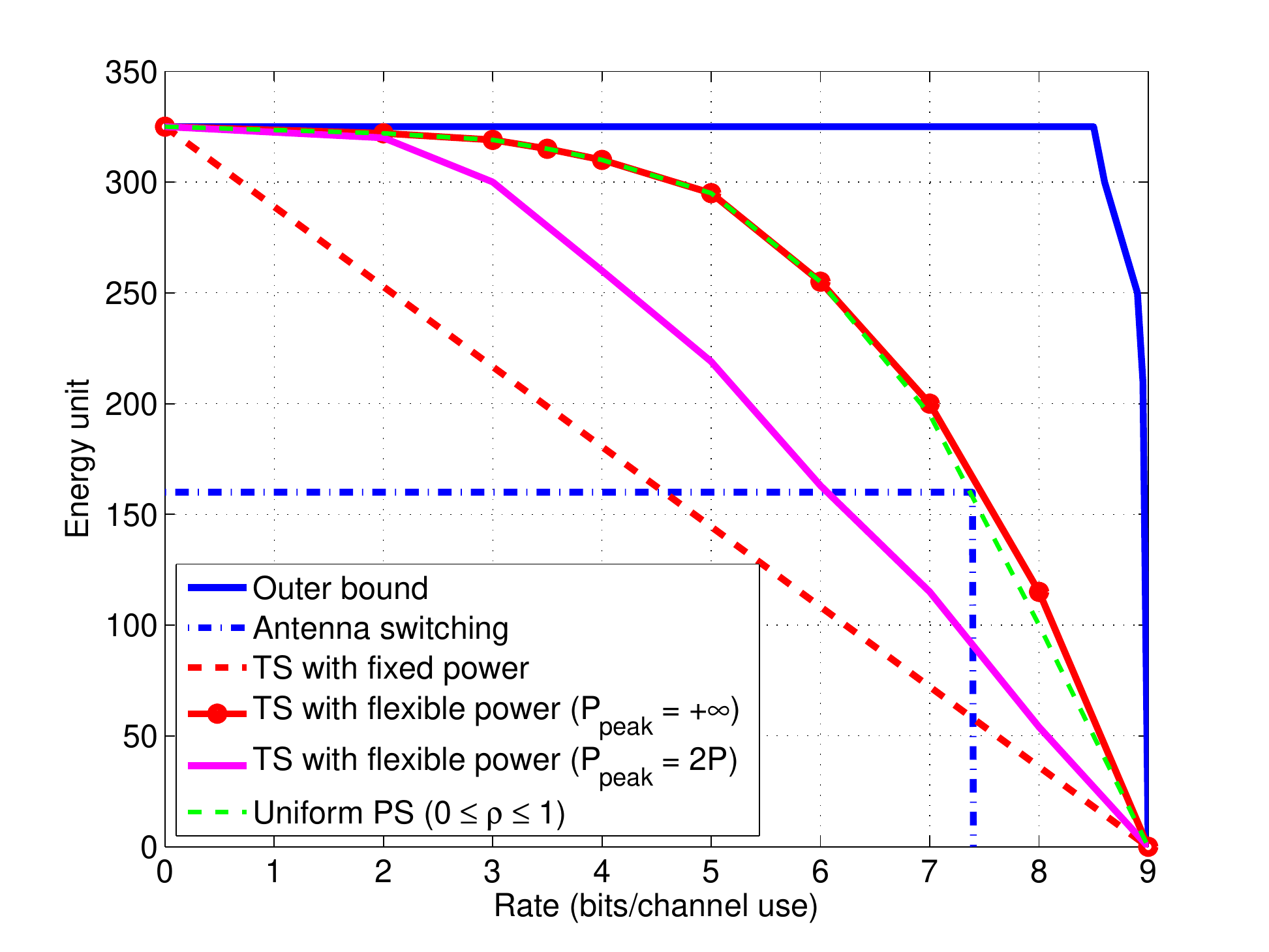}
\caption{The R-E region of a $2$-by-$2$ MIMO with a single receiver for both the energy and information, where
$\mathbf{H} =[1 ~0.8; 0.8~ 1]$ \cite[Fig. 8]{rui13}.}
\label{recolocate}
\end{figure}

Multi-antenna beamforming
{\red can potentially enhance the implementation and operation of SWIPT.
An MIMO wireless broadcast network with three nodes is studied in \cite{rui13},
where, apart from a source node, one of the nodes gathers RF energy and another node receives information.
All of the nodes can have multiple antennas.
Two interesting cases are investigated in \cite{rui13}.
In the first case, the above-mentioned two nodes are far apart.
They have distinctive MIMO channels from the source.
In the other case, the two nodes separately receiving energy and information are
co-located and therefore they have identical MIMO channels from the source.
In the first case,
the optimal transmit policy is derived to optimally trade off information data rate for energy delivery,
as can be portrayed by a so-termed rate-energy (R-E) region.}
Fig. \ref{reseparate} \cite[Fig. 4]{rui13} plots this R-E region,
where $M, N_{\text{EH}}$ and $N_{\text{ID}}$ are the number of antennas at the source node,
the node harvesting RF energy, and the node receiving information data, respectively,
$Q$ is the total harvested RF power, and the transmit power is $P=1$W.
{\red In the second case, the attainable R-E region is demonstrated,
which is potentially limited in practice due to the incapability of the EH node to decode information directly.
Restrained by this shortcoming, two practical operating policies}
are developed for the second scenario, referred to as time switching (TS) and power splitting (PS).
The attainable R-E regions of the policies are characterized in Fig. \ref{recolocate}  (i.e., \cite[Fig. 8]{rui13}),
where $P=100$.

Multiuser collaborative MIMO SWIPT networks are investigated in \cite{qin17}
{\red where non-negligible circuit power consumption is included.
In specific, the authors of \cite{qin17} aim to maximize the overall data rate of all the active users in the system uplink
while providing satisfactory QoS to the downlink service of the users.
The task is then transformed into a convex program.
The beamformers and operating time, and the PSFs of every user are optimally designed
with a joint consideration of both the uplink and downlink.
SDP and golden search are used for the optimal design.
Additionally, an optimally selected subset of users are activated in \cite{qin17}, which is achieved by first activating the users at the beginning
and then turning off those contributing negatively to the growth of the sum rate.
Numerical results indicate that the approach developed in \cite{qin17} incurs a lower computational complexity
at a marginal penalty of sum rate, in comparison with the conventional combinatorial integer programming alternative.}

The authors of \cite{khan}
{\red develop a malleable system structure to portray the performance}
of wireless power and information transfer enabled by an mmWave cellular network,
where devices can align their beams to that of the BS, or where no such beam alignment is undertaken.
{\red For the two operating scenarios, the authors investigate the function of several device-dependent}
parameters on the system performance.
The authors find out that
the total (power and information) coverage probability can be enhanced by optimizing the PSF
to optimally allocate the received signal between the EH and the information decoding segments.
{\red To utilize multiple antennas at the receiver with the least power consumption,}
they develop a simple switch-based receiver framework for SWIPT.

Many existing designs of SWIPT systems are based on Gaussian inputs,
{\red which may result in significant performance deterioration
when applied to the situation with finite-alphabet inputs.}
Authors of \cite{zhu17} design the precoder for such situation, in the presence of
real-time CSI of an MIMO channel.
They construct the precoder design task as an optimization problem
{\red to maximize the reciprocal information from the transmitter to the receivers,
constrained by the transmit power and a requirement on EH amount.
Since the problem is NP-hard, the global optimum cannot be obtained in a polynomial time.}
Utilizing its structure, they relax the problem to an SDP problem.
By applying the SDR technique,
they develop a general solver for both co-located and separated receivers to realize a near-optimal beamforming precoder.
In a special scenario where multiple receivers are co-located, the authors
demonstrate that the optimal design of the beamformer exhibits strong concavity in regards to the transmit power.
A specialized algorithm is developed particularly for this special scenario, and it
{\red demonstrates in \cite{zhu17} nearly identical effect yet with a substantially lower complexity,
as compared to the SDR solution designed for the general scenario.}

In \cite{xing2013}, the authors investigate MIMO wireless communication networks
{\red constrained by EH amounts.
The studied EH network includes
one transmitter, one receiver, and multiple EH nodes.
The EH nodes can convert their captured electromagnetic waves into power to
extend the system operating duration.}
When the transmitter sends data to the receiver, it should also optimize the beamforming/precoder matrix to charge the
{\red EH nodes in the meantime.
Additionally, the amount of the charged energy should exceed a given threshold.}
Under the EH constraints, both the minimum mean-square-error (MMSE) and reciprocal
{\red messages are considered as optimization metrics
to contrive the beamformers at the transmitter.
In order to make the developed approaches adaptable for real-world implementation with reasonable overhead,
the authors of \cite{xing2013} also generalize the beamforming policies with partial CSI.}

\begin{figure*}[th]
\centering
\includegraphics[width=0.7\textwidth]{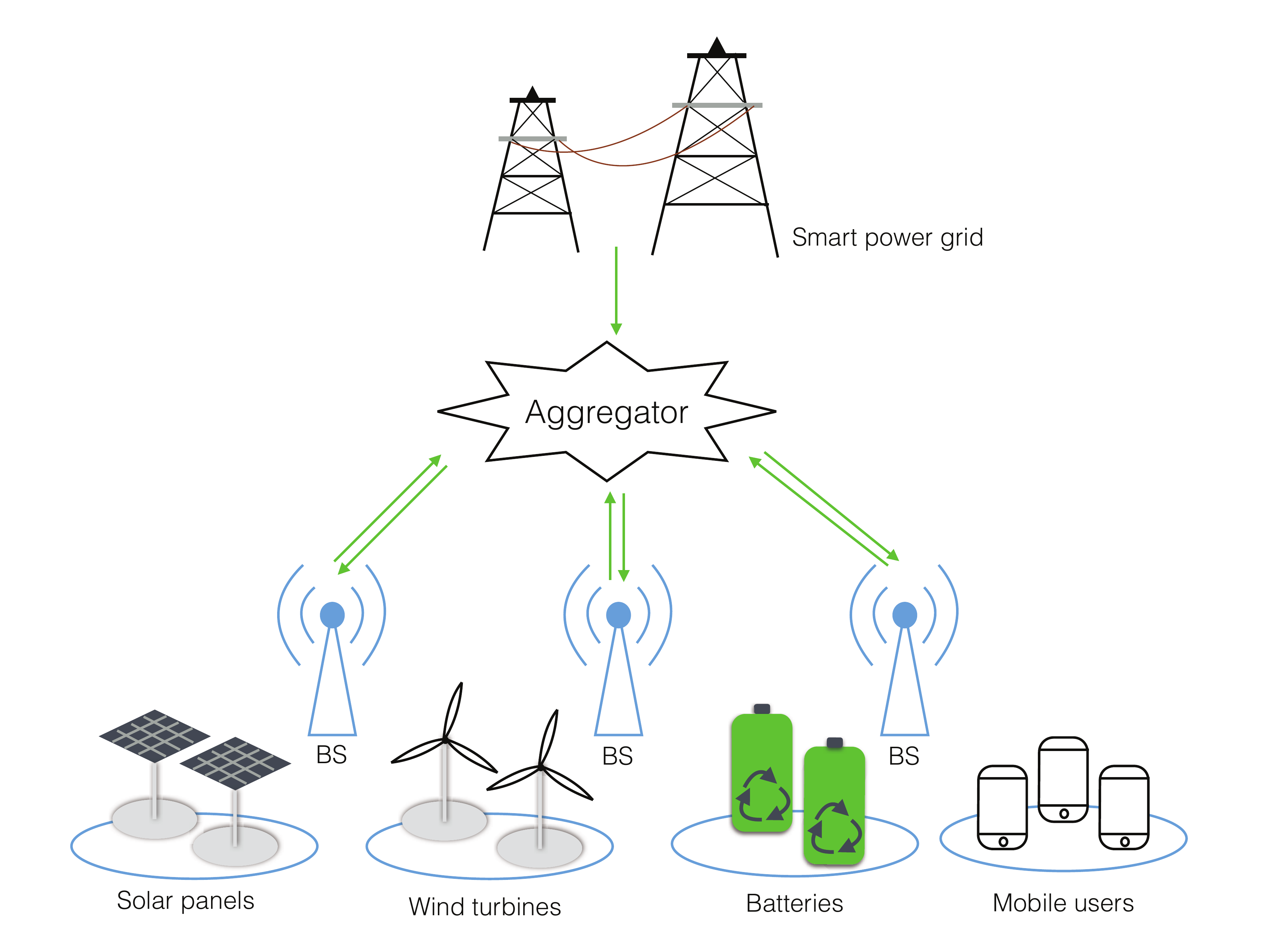}
\caption{A smart grid-powered wireless cellular network \cite{wang2015}, where each BS serves several mobile users and is equipped with RES devices and batteries. The BSs can carry out bidirectional energy transactions with the smart grid, and share energy with other BSs through the aggregator.}
\label{system}
\end{figure*}

\section{Smart Grid-Powered Wireless Networks}\label{sec.sharing}

The traditional power grid is shifting to a ``smart'' one with many state-of-the-art new functionalities, such as smart metering, RES, demand-side management, dynamic pricing, and two-way energy trading. Powered by such an electricity grid with one or more emerging techniques, cellular networks can have more choices in energy types and can design energy-efficient operation schemes.
For example, there can be redundant energy at some BSs, which can be redistributed to power other BSs and their services.
Fig. \ref{system} depicts a typical framework of the smart-grid powered cellular networks,
where each BS serves several mobile users and is equipped with RES devices
(such as solar-electric converters and/or hydroelectric generators) and batteries.
The BSs can carry out redistribution of unused and/or redundant energy to enhance the EE across the entire systems.

Featuring (part of) the system model in Fig. \ref{system}, many recent works focus on enhancing system performances by
minimizing energy consumption and operational cost \cite{yeow14, gong14, che16, sheng17, hu17con, hu17, farooq, Xu15},
or maximizing system's efficiency and operator's utility \cite{ghaz17, ghaz, zhang17, rama}.
State-of-the-art techniques and methodologies, such as game theory,
DP \cite{dp1966, dp78, dp95}, and stochastic gradient descent (SGD) \cite{sgd10, sgd13, sgd15, sgd16}, can be leveraged.
The objectives and the solving techniques of the optimization problems are categorized and summarized in Table \ref{tab.trade}.
Game theory is applied when competition or cooperation exists in the system consisting of BSs and electricity retailers.
{\red DP usually decomposes a sophisticated problem into a set of much easier subproblems with recurrence expressions.}
This relationship is known as Bellman optimality equation.
SGD is usually applied to achieve the asymptotically minimum time-average expenditure of a network
in no need of any prior knowledge
of system's randomness (such as the EH amounts and electricity prices).
The real gradient value of the objective is replaced with the gradient from the training set,
which helps to decouple the optimization and constraints over time.
It is proven that the optimality loss of the objective can asymptotically diminish by reducing the stepsize of SGD~\cite{sgdiot}.
The application scenarios, merits and disadvantages of some popular algorithms are summarized in Table \ref{tab.alg}.
Conditions and auxiliary methods applied to solve the problems are summarized in Table \ref{tab:aux}.

In Sections \ref{sec.sharing}, \ref{sec.purchasing}, and \ref{sec.trading}, we will review the literature on smart grid-powered cellular networks with different functionalities. We first focus on the RES powered systems in Section \ref{sec.sharing} and discuss the works on energy harvesting and sharing.
Then we steer the survey to smart-grid powered systems in Section \ref{sec.purchasing} with energy purchasing based on dynamic prices.
Finally, we review the works on two-way energy trading in Section \ref{sec.trading}.

\begin{table*}[t]

\renewcommand{\arraystretch}{1.4}
\centering
\caption{Optimization objectives and methods for smart grid powered communication systems}
\label{tab.trade}
    \begin{tabular}{ | p{3.5cm}<{\centering} | p{1.8cm}<{\centering} | c | p{1.6cm}<{\centering} | c | p{0.8cm}<{\centering}  | c | p{1.6cm}<{\centering} |}
    \hline
    \diagbox{\makecell{Optimization \\ methods}}{\makecell{Optimization \\ objectives}}

    &\makecell{Total energy \\consumption}    &\makecell{Operational \\ cost}     &\makecell{Utility of \\ operators}
    &\makecell{Energy/\\cost \\ efficiency}    &\makecell{QoS}     &\makecell{Weighted/ \\ expected \\ sum rate}
     &\makecell{Long-term \\ average cost}    \\ \hline

    Online greedy/heuristic algorithm      &\cite{yeow14, liu15}    &     &      &      &       &        & \\ \hline

    Stochastic sub-gradient method      &    &\cite{Xujie15}     &      &      &       &\cite{wang2016}        &\cite{wang2016dyn}  \\ \hline

    Lyapunov-based online algorithm     &    &     &      &      &       &    &\cite{wang2016two, xiaojing2017, mao15, zhai15, dong17} \\ \hline

    Stochastic or deterministic ADMM approach   &    &\cite{hu17con, hu17}     &      &\cite{zhang17}      &       &      &  \\ \hline

    Lagrange dual based (sub-gradient) method     &\cite{wang2015, shan16}    &\cite{Xu16}     &      &\cite{xures}   &      &\cite{hu16con, hu16}              &  \\ \hline

    Multi-stage or nested optimization     &\cite{gong13, gong14, farooq, liu15}      &\cite{Xujie15}      &\cite{Bu12, bu122, bu13}      &      &       &\cite{hu16}          &     \\ \hline

    Ellipsoid method      &    &\cite{Xu16}     &      &      &       &          &  \\ \hline

    Evolutionary algorithm      &    &     &\cite{ghaz}      &      &       &           &  \\ \hline

    Iterative algorithm      &\cite{sheng17, liu15}    &     &\cite{ghaz17, li2016}      &\cite{rama}      &       &          &  \\ \hline

    Dynamic programming      &\cite{gong14}    &\cite{che16}     &      &      &       &        &  \\ \hline

    Game-theory based method    &   &\cite{Xujie15}     &\cite{bu122, bu13}      &      & \cite{Bu12, ghaz14, li2016}      &          &  \\
    \hline

\end{tabular}
\end{table*}

\begin{table*}[t]
\renewcommand{\arraystretch}{1.5}
\centering
\caption{Analysis of algorithms}
\label{tab.alg}
    \begin{tabular}{ | p{3.5cm}<{\centering} | p{3.5cm}<{\centering} | p{3.5cm}<{\centering} | p{3.5cm}<{\centering} |}
    \hline
Algorithms     &Application scenarios     &Merits      &Disadvantages    \\ \hline
Greedy algorithm \cite{barron08, yeow14, liu15}     & A wide range of problems         & Obtaining local optimum in a fast and efficient way
                                                                       & Seldom achieving global optimum             \\ \hline
Deterministic ADMM  \cite{zhang17}              & Distributed computation         & Fast convergence        & Many samples needed per iteration to deal with stochasticity              \\ \hline
Stochastic ADMM   \cite{hu17con, hu17, ouyang2013}   & Stochastic and distributed computation         & One sample per iteration         & Converging with oscillations             \\ \hline
Lagrange dual-based sub-gradient method  \cite{wang2015, Xu16, xures, hu16con, hu16}     & Non-differentiable functions         & Converging to global optimum for (pseudo)convex functions under certain conditions       &    Only locally optimal for non-(pseudo)convex functions; iterative computation           \\ \hline
Stochastic sub-gradient method   \cite{Xujie15, wang2016, wang2016dyn}      & Time-coupling sub-gradients; sum-minimization problems where certain parameters are to be estimated & Smoother convergence; time-decoupling    & Near-optimal; iterative computation          \\ \hline
Lyapunov-based online algorithm \cite{wang2016two, xiaojing2017, mao15, zhai15, dong17}     & Time-coupling long-term scheduling horizon; uncertain models          &Time-decoupling; ensuring system stability and system penalty minimization         & Near-optimal; iterative computation                \\ \hline

\end{tabular}
\end{table*}

\begin{table*}[t]
\renewcommand{\arraystretch}{1.5}
\centering
\caption{Conditions and auxiliary methods}\label{tab:aux}
    \begin{tabular}{ | c | c |}
    \hline

Conditions and auxiliary  methods       & References                    \\ \hline
S-lemma                                              &  \cite{wang2015}                   \\ \hline
Semi-definite relaxation technique      &   \cite{wang2015}                 \\ \hline
Time-sharing condition                         &  \cite{ng13}                  \\ \hline
KKT optimality condition                       & \cite{xures, shan16}                  \\ \hline
Uplink-downlink duality                         &  \cite{Xu16}                  \\ \hline
Zero-forcing beamforming                    &   \cite{Xu16, dong17}                  \\ \hline
Game theory (Nash bargaining, Stackelberg game )       &\cite{Bu12, ghaz14, bu122, bu13, li2016}          \\ \hline

\end{tabular}
\end{table*}


\subsection{Utility Maximization} 

Ramamonjison et al. \cite{rama} study the resource allocation problem for a two-tier wireless system,
where smart grid-powered BSs can share renewable energy and battery storage through the aggregator.
The authors aim to maximize the system EE
{\red while meeting the demand of an average sum-rate at each cell.}
They first design an extended convex-concave procedure to tackle the non-convexity issue in the problem,
and then take a well-known Dinkelbach method \cite{Ished} to address the resultant subproblems.
The offline algorithms developed in \cite{rama} have a polynomial-time complexity
as the inner optimization problems can be
{\red resolved in a polynomial time by standard convex solvers like the interior point method~\cite{boyd}.}

The authors of \cite{zeng2014} and \cite{zeng2016} design multi-antenna beamformer for EH transmitters
according to finite-alphabet inputs and the statistical knowledge of the CSI at the transmitters in real-time.
{\red They aim to maximize
the sum of the average reciprocal messages within a channel frame without violating the causality of the EH process.
This results in a $2N_t^2$-dimensional stochastic DP problem.
The objective of the problem exhibits non-concavity
in which $N_t$ is the number of transmit antennas.
The authors of \cite{zeng2014} and \cite{zeng2016} prove the equivalence of the multi-dimensional stochastic DP problem
to a one-dimensional problem to select the transmit power level.
Dealing with the one-dimensional alternative can alleviate the computations without penalizing the optimality.
The one-dimensional alternative is first interpreted as
a discrete-battery-state discrete-power-choice task and solved by backward recursion \cite{Bert05};
and then translated to a continuous-battery-state continuous-power-choice task
and tackled by approximating the one-dimensional objective over a continuous feasible solution region.}

\subsection{User-Engaged Energy-Efficient Communication Scheme}

{\red A resource allocation task is studied in \cite{xures} for energy-collaboration}
enabled two-tier heterogeneous networks (HetNets) with NOMA,
including a macro BS and several pico BSs.
User association and power control are optimally designed to maximize the efficiency in the energy utilization of
the whole system under QoS constraints.
To achieve this, a decentralized technique is first proposed to
{\red draw the optimal user association given a transmit power.
Then, user association and power control are jointly and optimally specified.
This allows for far higher EE than other alternative policies, such as those developed in \cite{saito13, ding16}.
}

A network is proposed in \cite{niyato16} where mobile end users can share energy with each other at their encounter,
hence minimizing the
{\red probabilities of inadequate energy for their consumption.
Optimization of the corresponding network contains two major stages.
The first stage shares energy optimally amongst the mobile end users}
who agree to share (in other words, a couple of matched users).
{\red A stochastic optimization problem is constructed to acquire this optimal scheme,}
by considering the mobility patterns and the energy availability of the users.
The second stage of the developed network is to design 
a steadfast user-matching policy.
{\red Each individual of the users finds a peer as its partner to share energy.
The scheme drawn in the earlier stage can be utilized in such a way that a pair of well matched users
are most unlikely to undergo an energy outage.}

Energy-aware traffic offloading schemes are studied in~\cite{shan16},
where small BSs (SBSs) are powered by conventional electricity grid and/or RES.
User associations, on/off modes of SBSs, and power control are
{\red collectively optimized based on
the statistical data of energy and traffic arrival.
In a single SBS scenario, the closed-form expression for energy saving gain is derived,
which facilitates the calculation of the SBS activation and traffic offloading strategy.
A two-phase energy-aware traffic offloading approach is further developed in a multiple-SBS scenario,
taking into account multiple characteristics of SBSs with a range of energy sources.}

A user association problem is first formulated in \cite{liu15} via convex optimization in the space dimension.
Total energy consumption is minimized by
{\red allocating the traffic among different BSs
dynamically in a given time slot.
Then, RES allocation is optimized over different time slots for each BS to minimize the usage of the
energy from the grid.
To tackle this optimization task, a low-complexity offline approach with infinite battery capacity is designed
by assuming non-causal RES amounts and data traffic statistics.
The offline algorithm can achieve the optimality in this scenario, and play the role of a
performance upper bound for evaluating practical online approaches.
Heuristic online approaches with finite battery capacity are further developed
which are reliant only on causal RES and traffic information.}

\subsection{Minimization of Energy Consumption}
Chia et al. \cite{yeow14} present a new model to describe energy cooperation among BSs powered by a smart grid
(including conventional electricity grid and RES), limited energy storage, and connection by resistive power lines
to share surplus electricity.
They aim to minimize the expectation of the electrical energy supplied by the conventional grid and utilized by the BSs, i.e.
\begin{equation}\label{ce}
\mathbb{E}_{E_i^t}\big[\frac{1}{T}\sum_{t=1}^T(w_1^t+w_2^t)\big],
\end{equation}
{\red where $E_i^t$ is the EH amount captured by BS $i$ at time $t$,
$w_i^t$ stands for the energy procured from the traditional electricity grid,}
and $T$ is the total scheduling time horizon.
If the renewable energy profile (REP) and energy demand profile (RDP) of all the BSs
{\red are deterministic or known in advance (Case 1),
the optimal energy collaboration scheme for BSs can be awarded straightforwardly by tackling a relatively simple linear programming problem,
since the objective function in \eqref{ce} is simplified linearly to be}
$\sum_{t=1}^T(w_1^t+w_2^t)$,
and all the constraints of \eqref{ce} are linear in the first place.
If the REP and RDP are two stochastic processes and only causally known at the BSs (Case 2),
Chia et al. propose a greedy online algorithm by {\red taking a quick picture of the aforementioned linear optimization problems}
(Case 1) with $T=1$.

Let $r_{\alpha}$ and $r_{\beta}$ be the energy loss coefficients
which describe the ratio of loss during charging the battery and transferring energy between BSs, respectively.
Chia et al. \cite{yeow14} analyze the optimal structural properties of the greedy online algorithm under certain conditions.
For example, if $r_{\beta}=0$ or $r_{\beta}=1$, the greedy method can preserve optimality for any feasible energy profiles.
If $r_{\beta} \ge r_{\alpha}$, then energy transfer is optimal.
That is, if $E_1^t \ge 0 \ge E_2^t$ at time slot $t$, then sending $\Delta = \min \{|E_2^t|/r_{\beta}, E_1^t\}$
units of electricity from BS $1$ to BS $2$ to {\red make up for the shortage of $E_2^t$ is part of an optimal scheme.}
If $E_1^t \geq 0, \forall t$ and $r_{\beta} \ge r_{\alpha} $, the greedy policy is optimal.
According to the symmetry between $E_1^t$ and $E_2^t$, the same result holds if $E_2^t \geq 0, \forall t$ and $r_{\beta} \ge r_{\alpha}$.

Chia et al. \cite{yeow14} also provide insightful analysis and conclusions:
i) for the optimal scheme, no electrical energy should be bought to increase the battery levels from the grid;
ii) a system beginning with highly charged batteries can have a low optimal cost;
and iii) it is more cost-saving to save energy locally at each BS than to transfer and save the energy at the other BS.
On the other hand, it is reasonable to assume partial knowledge of REP, which consists of
a deterministic waveform with small random noises added at each time slot as the prediction errors.
Inspired by the aforementioned two cases, the authors of \cite{yeow14} then propose a
{\red compound approach which can utilize offline statistics
about the REP and operate in an online fashion.
They use the non-real-time offline approach to draw the adequate policy (or schedule) for the deterministic part of the
REP, and then apply the greedy heuristic to recoup any gaps pertaining to non-negligible noise effects.}

The average grid power consumption is minimized in \cite{gong14} for RES-powered BSs
under users' QoS (blocking probability) constraints.
The task is converted into an unconstrained optimization problem to minimize
the weighted sum of the grid energy usage and blocking probability.
{\red A two-phase DP approach is developed
by leveraging statistical data for traffic load and RES.
The BSs' on/off
modes are optimally decided in the first phase.}
The active BSs' resource blocks are assigned iteratively in the second phase.
Compared with the optimal collective BSs' on/off modes and active resource blocks allocation approach,
{\red the proposed approach significantly decreases
the computational complexity
and can realize the optimal operation when the traffic obeys a uniform distribution.}

The grid energy expenditure is minimized for smart grid-powered cellular networks in \cite{sheng17}.
The task is formulated as an NP-complete mixed-integer nonlinear program.
For centralized systems, a cost-aware approach is designed to tackle the load distribution problem
and the energy configuration problem in an alternating manner.
The centralized algorithm requires a low computational complexity, and rapidly converges to the near-optimal solutions.
For distributed networks,
{\red a three-stage decentralized ruling scheme is proposed in \cite{sheng17} where the BSs and mobile terminals can independently calibrate
their individual policies only based on limited knowledge locally accessible to them.}
System expenditure can be greatly reduced in both types of networks.


\subsection{Lyapunov-based Online Optimization}
The grid energy consumption (GEC) is minimized in \cite{zhai15} for a smart grid-powered
queueing orthogonal frequency-division multiple-access (OFDMA) system with battery leakage.
{\red By considering the temporal variation of the network, the GEC minimization task is formulated
to collaboratively design the admission regulating, power assignment,
subcarrier distribution, and communication duration.
The random RES amounts are simplified to be an i.i.d. process.
By taking advantage of the state-of-the-art Lyapunov optimization, an efficient online approach is developed,
termed as leakage-aware dynamic resource allocation strategy.
This strategy tracks present system states which consist of CSI and battery level}
with no need for the a-priori knowledge on the states.
Furthermore, it is proven that the minimum GEC value can be
{\red reached asymptotically with the proposed approach.}

The network service cost is introduced in \cite{mao15} as a performance metric
to account for both the grid
{\red power usage and attainable QoS.}
A computationally inexpensive online approach is proposed to minimize the average system service expenditure
over a long time by conducting joint BS association and power control (BAPC),
referred to as the Lyapunov optimization based BAPC (LBAPC) algorithm.
{\red A merit of the approach is that the decisions rely solely on the instantaneous state knowledge with no need for any
a-priori knowledge on the distribution statistics of CSI and EH processes.
To decide the system activity,
only a deterministic problem needs to be solved at every time slot.
A simple inner-outer optimization approach is developed to provide effective solutions to the deterministic problem.}
It is further proven that the LBAPC approach is asymptotically optimal,
as the control parameter $V$ (with unit J$^2$/cost) $\rightarrow \infty$.

\section{Energy Planning Under Dynamic Pricing}\label{sec.purchasing}

Dynamic pricing is a vital mechanism of smart power grids.
It can help shift some load in peak time to off-peak time,
and thus balance the power consumption in these two periods for the power grid.
Cooperative BSs can plan their energy purchasing, storing, and sharing jointly according to the prices
to reduce the system-wide operational expenditure.

\subsection{Operational Cost Minimization}

In \cite{che16}, the on-grid energy expenditure is minimized in a large-scale green cellular system
{\red by collectively designing the optimal BS on/off strategy and electricity procuring scheme.}
The green cellular network often experiences fluctuations brought by RES, energy prices,
{\red wireless data traffic, BS coordination, and traffic offloads.
Hence, it is usually  NP-hard to obtain the optimal scheme that minimizes the electricity
expenditure over a long-term meanwhile guaranteeing the users' QoS.}
For a dynamic system design, stochastic geometry (Geo) can be applied to account for
large-scale wireless network.
DP can be used to develop adaptive on/off schedule of the BS
and the electricity procurement.
{\red By integrating these two aspects, a new
Geo-DP design is shown to guarantee that the optimal probability of the BS being activated (or staying ``on'')
can just suffice the QoS requested by the users.
Nonetheless, the typical curse of dimensionality of DP \cite{dp1966}
prevents the optimal electricity procurement scheme from scaling up to large-scale wireless networks.
Therefore, a suboptimal energy procurement scheme with a low complexity is put forth,
where on-grid electricity is procured in abundance with an adequate price only if the present battery reading
and the expected future RES level are both low.}
This suboptimal policy can realize a near-optimal performance.

\subsection{Utility Maximization}

The authors of \cite{ghaz} aim to maximize the profit of
cellular operators meanwhile minimizing the CO$_2$ emissions in green cellular networks
and satisfying the desired QoS.
The cellular network is powered by the smart grid where electricity retailers sell renewable energy to the BSs.
Energy from different retailers may have various types, and thus have distinctive prices and pollutant levels.
The BSs' sleeping policy and electricity purchase policy are decided via several algorithms,
such as an iterative algorithm and an evolutionary algorithm
(including the genetic algorithm and the particle swarm optimization method).
For the iterative algorithm, it is assumed that all BSs are switched on at the beginning.
{\red Then, at each precluding stage, one BS is shut down at a time.
For the $i$-th precluded BS, the corresponding optimal utility function is calculated
and compared with the previous maximum utility to determine whether precluding BSs is possible or not.
The algorithm terminates when there are no more BSs to be precluded.}

The utility of the mobile operators is maximized in \cite{ghaz17}, which collaboratively make decisions on
energy procurement and BS sleeping policy based on profits,
network demands, retailers' capacity, and CO$_2$ emissions.
Energy harvesting, dynamic pricing, and energy sharing are implemented in the framework
to help reduce energy consumption of the network.
Three utility metrics are introduced to
{\red measure the level of fairness}
in optimization, including the weighted sum, the max-min criterion, and proportional fairness (PF).
The sum utility metric measures and quantifies the overall profit of the network.
This metric promotes operators with higher revenues, while depriving the possibility for low-revenue operators
to purchase the cheapest energy.
The max-min metric maximizes the minimum profit across the network, thus improves fairness to the system.
The PF metric maximizes the geometric mean of the profits, which can efficiently avoid very low profit since
a close-to-zero profit would
{\red cause the entire objective function to diminish.}
Given the BSs' on-off state, the energy procurement problem exhibits convexity and can be readily tackled by the Lagrangian method.
Given the optimal energy procurement, the BS on-off switching problem is non-convex.
A deterministic iterative approach is first developed to establish the optimal on-off policy for the BS in a centralized fashion.
Later, a decentralized algorithm is proposed with faster convergence and lower complexity,
yet penalizing performance gain.

\subsection{Game Theory for Energy Procurement}
Liberalization of the electric power market has been advocated in many countries and regions \cite{Bu12},
where electricity retailers can set their prices and compete for best interest.
To achieve energy-efficient green communication,
it is necessary to consider data traffic, dynamic electricity price, the pollutant level {\red incurred by brown energy} consumption,
and the robustness of the smart grid when shaping green wireless cellular mobile systems.
Every BS is expected to choose its most cost-effective electricity retailers at any moment
for economical and ecological profits \cite{Bu12}.

Aligned with this goal, the smart-grid powered cellular network system is constructed
as a two-level Stackelberg game in \cite{Bu12, ghaz14}.
At the cellular network level,
the active BSs can choose the electricity retailers and the corresponding purchasing amounts,
aiming to achieve the lowest service blocking probability with the least possible expenditures.
At the smart grid level,
the retailers can decide on their electricity prices such that they can acquire as much extra profit as possible,
competing to get selected by the BSs at the same time.
Based on the Lagrange dual method, the existence and uniqueness of the Stackelberg equilibrium
are proven in \cite{Bu12} for the proposed Stackelberg game;
while an iterative algorithm with low complexity is applied in \cite{ghaz14} to draw the optimal solutions.
It is shown that the smart grid has substantial influence on green wireless cellular systems,
{\red and the developed strategy can greatly diminish the energy expenditure}
and $\text{CO}_2$ emissions in these systems.

The same mechanisms developed in \cite{Bu12} and \cite{ghaz14} are extended to
a three-level Stackelberg game for cognitive HetNets in \cite{bu122, bu13},
where the three parities in the game are smart grid retailers, macro-cell BSs (MBSs), and femto-cell BSs (FBSs).
A homogeneous Bertrand game is designed to model the price decisions made by the retailers.
Then, a backward induction approach is leveraged to analyze the designed game.
It is observed that both the FBSs and the MBSs intend to choose the electricity retailer with the cheapest price.
The electricity prices are set according to the Nash equilibrium at the smart grid level.
In this case, no retailer can increase its individual net
{\red profit by selecting a different price,
given the prices provided by the other retailers.}

The RES is utilized to power millimeter-wave (mmWave) backhaul networks in \cite{li2016},
{\red where wireless operators purchase electricity from several renewable power suppliers
to support mobile terminals.
A lead time-dependent pricing strategy is developed, which enables a wireless operator to control
the latency of traffic and service deliveries over the backhaul links and coordinate the suppliers to decide on
how much RES to be stored at each supplier.
The task is constructed as a standard Stackelberg game between the network operator and several renewable power suppliers.
Different from \cite{Bu12, ghaz14, bu122, bu13},
the wireless operator acts as the leader in this game, who determines the pricing mechanism for the power suppliers,
while the suppliers play the followers who decide their self energy storage policies according to any given pricing scheme.}

In a centralized network, the operator and the renewable power suppliers jointly maximize the system profit \cite{li2016},
while in a decentralized one, the operator and the renewable power suppliers maximize their individual profit.
{\red Effective decentralized methods are developed in \cite{li2016} to acquire the optimal pricing strategy of the wireless operator,
and the Pareto equilibrium storage policies for the suppliers, respectively.}
Both the algorithms start by treating each renewable power supplier as a separate group.
For the operator's game, the algorithm merges two neighboring groups in each iteration.
For the suppliers' game, the algorithm decides whether or not to merge two groups
by comparing their supportable traffic loads in each iteration.
Both algorithms stop when no more mergers happen.
{\red It is also shown in \cite{li2016} that the proposed distributed policy allows a wireless operator to generate a}
higher profit than a centralized scheme.

{\red Traditional electricity market is shifting to a ``smart'' one, offering several electricity purchase policies and their combinations
for mobile network operators (MNOs) of cellular networks.
It can be possible for the MNOs to prepay for electricity day-ahead
at a relatively economical price and also purchase electricity based on a real-time demand at
a relatively high and less economical price, respectively \cite{dms06}.
To reduce the power expenditure, it can be very important for MNOs to comprehensively organize}
their day-ahead and real-time electricity purchases {\red according to their dynamic traffic load over time.}
The BSs of the MNOs can offload their traffic and services, so as to switch the most number of lightly-loaded BSs
into the sleep state for system-wide energy-efficiency.
To this end, the authors of \cite{Xujie15} take two different MNOs {\red co-located in a region for an example.
The two MNOs interplay in both electricity procurement and traffic load balancing for operational expenditure reduction.}
The two MNOs can both intend to minimize their own electricity expenditure,
two-phase stochastic programming is adopted to draw the optimal policy for electricity group purchasing with load balance.

In the case where the two aforementioned MNOs are from two competitive organizations
and are only interested in minimizing their own electricity expenditures,
the authors of \cite{Xujie15} develop a repeated Nash bargaining scheme,
where the MNOs bargain and split electricity expenditures
under electricity group purchasing and load sharing.
At the first stage, the two MNOs settle the deals on the day-ahead electricity group purchase,
as well as how to {\red split the collective electricity} commitments between them,
by considering the potential real-time collaboration benefits.
At the second stage, under given day-ahead electricity commitment and the corresponding electricity group buying,
the two MNOs negotiate in real-time at each slot $t$ about the real-time electricity group purchasing,
and how to {\red allocate the collective electricity} trading amounts, the wireless loads, and the inter-MNO payment.
This Nash bargaining scheme can {\red realize Pareto-optimality and,
in turn, the effective electricity expenditure decreases for both MNOs.}

\section{Two-way Energy Trading and Cooperation}\label{sec.trading}

The electricity bills of cellular operators continue rising due to explosive demands for wireless services
in the 5G and beyond communication networks.
The BSs are seeking new approaches to diminish the energy expenditure with the integration of RES.
The bidirectional electricity trading capability of smart power grids allows the cellular BSs
to sell their redundant renewable energy to the grid for profit,
which is a straightforward way to save the operational costs.

\subsection{Energy-efficient Schemes}
With the spatial variations of user density, data traffic, and the amount of RES,
it is more efficient for the BSs to collaborate and jointly schedule their energy, wireless resources, and/or downlink users.
An energy-efficient framework is proposed in~\cite{Xu15} where different BSs
share or trade electricity with the assistance of an aggregator in a smart power grid,
and/or share radio resources and offload traffic {\red to cut off the operating expenditure.}
The following three approaches are presented to achieve this goal.

The first approach is energy collaboration on the side of power supply.
The BSs use a bidirectional flow to merchandise redundant electricity with the smart grid or share renewable energy with each other
through the aggregator.
The two BSs in the same group schedule the energy to be injected to, or drawn from the grid simultaneously,
since their energy surplus and deficit can be matched.
The energy demands for communications are specified and analyzed.

The second approach is communication {\red collaboration} on the demand side.
{\red The BSs perform cost-oriented wireless communication collaboration to share resources and
reschedule traffic load both spacially and temporally,
by considering a preset amount of the energy (RES and/or traditional).}
Three different cost-oriented schemes are typically studied for various time scales to achieve such cooperation.

The first cost-oriented scheme is traffic offloading \cite{niu2010cell},
where the BSs {\red in short of RES can transfer their users to adjacent BSs
with ample RES (even if they bear heavier traffic}), to reduce the total amount of electricity purchased from the smart power grid in an attempt to save the operating expenditure.
Traffic offloading can be executed on a basis of several seconds.
The second cost-oriented scheme is spectrum sharing \cite{golds2009},
where BS$1$ shares {\red a portion} of its available spectrum with BS$2$.
Under the same users' QoS constraints, BS$2$ can reduce its transmit power procured from the grid,
while BS1 consumes more RES for transmission. Hence, the total cost is diminished.
This can be implemented at a time scale of minutes.
The last cost-oriented scheme to achieve cooperation is CoMP \cite{gesb2010},
{\red which helps pair the BSs' transmit power with their EH amounts}
by accommodating the BSs' transmit signals.
In particular, the BSs with larger RES amounts are expected to use higher transmit powers to provide stronger wireless signals to the users.
CoMP runs at a symbol- or frame-level on a typical basis of microseconds to milliseconds.
This can be much complicated but save more energy budget,
as compared to the two earlier schemes developed in \cite{niu2010cell} and \cite{golds2009}.

The third approach to achieving bidirectional energy trading and sharing is joint cooperation of energy and communication
from both the sides of demand and supply.
The BSs collaborate to {\red reduce their operating expenditure} (i.e. electricity bills) to the greatest extent.
A small cell network (SCN) with EH capabilities is studied in \cite{mao215}, from the perspectives of
outage probability, system performance, grid power consumption, and cell association.
Analysis and simulation in \cite{mao215} reveal that the outage probability declines with the density
of the BSs deployed in the SCN (denoted as $\lambda_{BS}$),
and that the grid power consumption $P_G$ also decreases with $\lambda_{BS}$.
To reduce $P_G$ or enhance system performance, it is more efficient to increase $\lambda_{BS}$ than it is to increase the EH rate.
When $\lambda_{BS}$ is extremely small or large, the battery capacity has little impact on $P_G$ or system performance.
As for cell association schemes,
the distance-based policy suffers performance loss as it overlooks the spatial variation of available energy at different BSs.
Therefore, conventional cell association strategies cannot be directly adopted in EH-SCNs.
It is demonstrated in  \cite{mao215} that the SNR-based scheme outperforms the distance-based scheme,
as it utilizes information on both the distance and the current energy state.
However, it still undergoes performance degradation as it makes decisions only {\red according to the present} system state
and overlooks the coupling in different transmission blocks among different users.

\subsection{Minimization of Operational Cost}
A transmit beamforming scheme is designed in \cite{hu17con, hu17} for a coordinated multicell system,
where the transmission is powered by a smart power grid.
Operating in a distributed fashion, the scheme uses the state-of-the-art stochastic alternating direction method of multipliers (ADMM) \cite{ouyang2013}.
The conditional-value-at-risk (CVaR) cost~\cite{rock2002, quar2008} is introduced to reduce the risk of extremely high cost of the system.
The long-term SINR is considered to guarantee users' QoS based on the stable downlink channel covariance matrices
$\mathbf{R}_{ijk}$, where $i,j \in [1, \ldots, I]$ are the BS indexes.
It is proven that when $\text{rank}(\mathbf{R}_{ijk})=1$ or when the number of BSs $I \leq 2$,
the optimal beamforming matrices $\{\mathbf{W}_{ik}^*\}$ of the SDR problem is always rank-one,
which means that the relaxation is tight, and the solutions $\{\mathbf{W}_{ik}^*\}$ are globally optimal.
Simulation results verify the tightness of the SDR in general cases.

The problem is solved offline in a distributed fashion via the stochastic ADMM without the {\em a priori} knowledge of the
EH amounts and electricity prices $\{s_i\}$.
The BSs update their primal and dual variables by visiting one sample of the historical realizations of $\{s_i\}$ at each iteration.
The historical realizations are stored in the data base of each BS $i$.
The BSs do not have to communicate their inter-BS interference powers with each other in this way,
which greatly reduces signaling and backhaul overheads.

\subsection{Utility Maximization}
{\red The idea of cost efficiency (CE) is applied in \cite{zhang17} to calculate the total data rate transmitted at the cost of a dollar}
for micro-grid (MG)-powered BSs.
The MG has smart grid features such as RES, bidirectional electricity trading, and dynamic price adjustment.
CE is maximized by jointly scheduling the electricity generation in the MG and optimizing the transmit power of the BSs.
The CE maximization task includes constraints accounting for the strong multivariate coupling over time.
To address this formulated fractional optimization problem, the Dinkelbach method is first applied.
Then a low-complexity method exploiting the ADMM approach is developed.
Several auxiliary variables are introduced to split variables into two sets, so that
the constraints capturing the different coupling are separable between the two subsets of all the optimization variables.
Consequently, the approach developed in \cite{zhang17} only accommodates simple updates at every step,
and thereby permits decentralized executions.
The approach is ensured to converge to the global optimum of original maximization of the CE.

Farooq et al. \cite{farooq} propose a mixed energy redistribution framework for mobile cellular networks,
which combines physical power lines and energy transaction between the BSs facilitated by the smart grid.
{\red According to the average value or the full statistics of the availability of the RES,}
algorithms are proposed to optimally deploy physical power lines between the BSs.
Taking into account the battery volumes and dynamic electricity pricing,
{\red a framework is
developed to schedule energy, and determine the optimal amounts of energy
and RES to be obtained and shared among the BSs, respectively.
Three cases are studied, in which the RES amount is unknown,
fully known, and partially known, in advance.}

\subsection{Lyapunov-based Online Optimization}\label{sub.lyap}

Based on \cite{wang2016two}, two-way trading and multi-timescale planning of energy in 5G networks are presented in \cite{xiaojing2017},
where implementation challenges are discussed, and the use of stochastic control theory is studied.
{\red The Lyapunov control theory is evaluated for potential applications.
The concrete usefulness of the theory can be validated by numerical simulation results in practical scenarios without
knowing the future CSI, transaction prices and EH amounts.}

{\red
The long-term energy cost minimization problem is studied in \cite{dong17}
for a smart grid powered communication network.
Facilitated by the two-way trading mechanism in the network, a harvest-use-trade strategy is selected
to deal with the shortcomings of the batteries
to enhance usage efficiency of the RES.
With the help of the Lyapunov optimization technique and the Lyapunov drift-plus-penalty function,
the stochastic optimization problem is reformulated into a joint minimization problem of energy
and packet rate.
Two suboptimal algorithms are proposed to tackle the NP-complete problem by considering
the CSI and the packet detection failure,
featuring the techniques of successive approximation beamforming (SABF)
and zero-forcing beamforming (ZFBF) \cite{wiesel2008}.
}

\section{Applications of Energy Harvesting and Smart Grid-Powered Wireless Communications to 5G/B5G}\label{sec.app}

The importance of energy harvesting and smart grid-powered wireless communications has also been found in many emerging applications in the 5G and beyond wireless networks,
such as mobile edge computing, machine learning, NOMA, URLLC, and so forth.

\subsection{Mobile Edge Computing (MEC)}
{\red Many recent and emerging developments of the IoT have opened up possibilities for a wide range of new mobile applications, requiring low-latency communication and intensive computation over massive mobile devices.
The emerging MEC has offered a new paradigm to offload computing workload from mobile users.
This can enhance computation capability for mobile users by leveraging remote execution,
and significantly reduce communication latency by having servers placed in the proximity of mobile users.}
By integrating the network function virtualization (NFV) techniques, MEC provides flexibility on the resource scheduling and service deployment~\cite{chen18m, chen19auto}.
Initial investigations~on~EH powered MEC systems have been conducted in
\cite{xu2016online,guo2018joint,zhang2018energy,zhang2018distributed,
mao2016dynamic,min2019learning,wei2018dynamic, feng17, lyu17},
focusing on EH-based MEC servers~\cite{xu2016online},~\cite{guo2018joint} and EH
mobile~users~\cite{zhang2018energy,zhang2018distributed, mao2016dynamic, min2019learning, wei2018dynamic, feng17}.~In~\cite{xu2016online, feng17}, the system operator learns online {\red the task amounts to be offloaded from the MEC server to the central cloud and the CPU frequency of the MEC server, based on the states of core network congestion and RES.}
Considering inter-cell interference in dense small cells, a distributed three-stage strategy is proposed in \cite{guo2018joint} to jointly
optimize task offloading, channel allocation, and computing resource allocation by decomposing the original problem.
Considering practical stochastic system environments, algorithms based on Lyapunov optimization techniques \cite{zhang2018energy,zhang2018distributed,mao2016dynamic} and deep learning \cite{min2019learning,wei2018dynamic}
have been used to produce dynamic task offloading and resource allocation strategies.

\subsection{Deep (Reinforcement) Learning}
Deep (reinforcement) learning has been widely employed in image processing and natural language processing \cite{krizhevsky2012imagenet,ren2015faster}.
{\red It has also been increasingly used to solve challenging
(in many cases,  non-linear non-convex) problems in wireless communication systems,
e.g., Polar decoding \cite{Gruber2017On, Cammerer2017Scaling}
and massive MIMO channel estimation \cite{He2018Deep, Wen2017Deep}.
Deep neural networks (DNNs) are able to solve sophisticated non-convex problems without explicit mathematical formulations~\cite{Sun2017Learning,Lee2018Deep,Wang2018A}.}
Several recent works have  investigated the EH-based wireless communication systems
where deep learning is utilized as an optimization method~\cite{do2019dynamic,kim2018action,chu2018reinforcement}.

\subsection{Non-Orthogonal Multiple Access (NOMA)}
{\red NOMA is a useful method to improve the spectral efficiency of wireless communication systems.
NOMA transmitters adopts superposition coding (SC) to stack the signals destined for multiple destinations or users.
Successive interference cancellation (SIC) is carried out at each of the destinations or users to extract the signals of interest
and suppress those intended for the others.
NOMA is able to connect}
multiple active users by using a single piece of time-frequency resources \cite{nomagsvd, nomaoppo},
and outperform its orthogonal counterpart in terms of spectral efficiency (SE) and flexibility.

EH-based NOMA is attracting increasing interests \cite{Zhang2017Performance,Min2018Energy,Alsaba18}.
In \cite{Zhang2017Performance}, a NOMA-based relaying network is
analyzed with a EH powered relay node to enhance system SE and user fairness.
{\red The outage performance of the network is examined, in which the transmit antenna selection is applied at the BS and maximal ratio combining (MRC) is applied at the end users. Closed-form formulations are established for the outage probability.}

Min and Meng \cite{Min2018Energy} study energy-efficient resource management for NOMA-based wireless powered sensor networks
{\red by constructing a EE maximization task regarding the EH time and transmit powers.
A particle-swarm-optimization-based approach is derived with fast convergence
by analyzing the structure of the optimization problem.
In~\cite{Alsaba18}, NOMA is first leveraged with beamforming to support several users in each beamforming vector.
SWIPT is applied to NOMA systems in \cite{nomarelay, nomasumrate, nomaswpt}.
It is proven in \cite{nomarelay, nomasumrate, nomaswpt} that integrating SWIPT not only facilitates cooperation among users,
but also alleviates the level of self-interference stemming from signal leakage from the output to the input.
Furthermore, the task of the total sum rate maximization is constructed
and tackled via a two-step convex programming based process.
The outage probabilities of both weak and strong users are established with closed-form solutions deduced.}

Some existing works have studied the power allocation on the multicast-unicast transmission \cite{nomapower},
spectral efficiency and confidentiality \cite{nomase} of NOMA,
as well as comparisons of system capacity between the emerging MIMO-NOMA and the existing MIMO-OMA~\cite{nomacapa}.
Transmit antenna selection at the BS is studied in~\cite{nomarelay}.
The weighted sum-rate is maximized for an SWIPT enabled cooperative NOMA system in \cite{nomasumrate},
by optimizing the power arrangement of the source and the
{\red PS coefficient at the users.
A new collaborative SWIPT NOMA network is developed in \cite{nomaswpt}, where the NOMA users near the source can
serve as EH relays to assist NOMA users far away.
Closed-form outage probability and sum throughput validate that
the application of SWIPT does not compromise the diversity gain, as compared with conventional NOMA.
It is confirmed in \cite{nomaswpt} that the opportunistic choice of node locations for picking users can realize a low outage probability
and yield better throughput than random picking schemes.}

\subsection{Ultra-Reliable Low Latency Communications (URLLC)}
5G/B5G mobile cellular networks are expected to support URLLC
which is expected to offer substantially short processing and transmission delays ($< 1$ ms),
meanwhile securing strong reliability (i.e. about $99.999\%$ successful delivery rate).
Resource allocation are studied
to enable URLLC in an OFDMA downlink in \cite{urharq, urmulti},
by minimizing the required system bandwidth or maximizing the system
{\red sum rate under QoS requirements.
An optimization problem is constructed}
in \cite{urpayload} to find the payload allocation weights that maximize the reliability
at targeted latency values.
A risk-aware machine learning approach is proposed for URLLC traffic management in \cite{urmachine},
by minimizing the risk of loss.
An SWIPT-enabled URLLC network is studied in~\cite{urswpt},
{\red where the credibility of the system is maximized by choosing the optimal SWIPT parameters
based on both the PS and TS protocols.}

\section{Lessons Learnt and Future Research Directions}\label{sec.future}
This article reviews the RES/smart-grid powered wireless communication systems, including
single point-to-point link, multi-hop link, multipoint-to-multipoint system, multi-point-to-point system, and multi-cell system.
In this section, we draw useful insights from the literature and explore promising future research directions.

\subsection{Point-to-Point Links}
As reported in Section \ref{sec.single}, one of the most popular techniques is the Lagrange multiplier method
which is applied to convex programming, as shown in Fig.~\ref{roadmap}.
The method has been utilized to unveil the underlying optimal structure of data transmission policies.
Several main challenges are worth future research.

\subsubsection{Learn-and-Adapt Algorithms for EH Systems}
Most existing works assume non-causal information about the EH, data arrivals and channel states
in the typical offline optimization framework, which is generally impractical. In the online scenario, the existing studies,
such as \cite{Xu2012Throughput,Chen2016Optimal} and \cite{gong13},
assume that there exists no a-priori knowledge on these system variations,
and develop heuristic schemes.
A potential online optimization technique is to employ learn-and-adapt algorithms \cite{chen17, chen19}
{\red which are expected to learn online the EH, data arrival processes and channel states,
and adapt the transmission strategy accordingly.
In \cite{blas}, Q-learning is considered to learn the optimal transmission policy with the EH,
data arrivals and channel states modeled as Markov (decision) processes.}
Exploring other learn-and-adapt algorithms with affordable complexity is an interesting research direction.

\subsubsection{Multi-User Networks and Energy Cooperation}

From a networking point-of-view, data transmission of multiple users/nodes (such as multi-hop and multiple-access networks) is of potential interest.
{\red The computational complexity of specifying the optimal data schedules increases generally with the user number.
Even in a two-hop link, the data schedule of the source can affect the data arrivals at the relay,
thus coupling strongly the transmission policy over the system \cite{Gunduz14design}.
Additionally, the causal information over different users is hard to obtain in practice.
Optimal schedules based only on current information need to be further investigated.}
Some basic multi-user scenarios have been studied in the literature, as summarized in Sections IV and V.


\subsection{Multipoint-to-Point Systems}
Apart from the data scheduling and energy management techniques for EH-powered multi-access networks
reviewed in Section \ref{sec.multiaccess},
potential future research directions are as follows.

\subsubsection{Power Allocation for EH-powered NOMA}
The power allocation policy for NOMA determines the interference cancellation capability of the receivers, and directly affects the throughput and user fairness of NOMA \cite{Dai2018A}. For EH-powered NOMA, power constraints need to be involved in the development of power allocation schemes.
The optimal scheme can be obtained by searching through the entire legitimate solutions (satisfying the power constraints),
which would lead to an excessively high complexity.
Low-complexity and dynamic power allocations constitute a promising research topic for future work.

\subsubsection{Channel Estimation and EH Prediction}
Most works on multi-access networks assume perfect CSI and non-causal information about EH, which is hardly possible in practice.
The design of practical channel estimators for NOMA is studied in \cite{Tan2016Novel} and \cite{Struminsky2016A},
and optimal approaches have been proposed to reduce the channel estimation errors.
However, the increase of the user number in 5G/B5G systems is expected to result in severe inter-user interference and,
in turn, severe channel estimation errors.
Moreover, few works have considered EH prediction for multi-access systems.
To this end, advanced channel estimation and EH prediction methods are required for multi-access systems.

\subsection{Multipoint-to-Multipoint Systems}
Existing studies on RES powered CoMP systems typically amount to minimize the total energy or cost.
Lagrangian dual-based methods and Lyapunov optimization methods are two popular classes of techniques used.
The optimality can be often achieved for convex problems, while asymptotic optimality is typically possible for online optimizations.
Some further research effort could be helpful to further advance the following aspects.

\subsubsection{Implementation of Uplink CoMP Techniques}
Most existing works study the joint transmission and energy management in RES powered downlink CoMP systems.
However, in practical network implementations, CoMP also supports
{\red accentuations to uplink reference information,
power regulating,}
and signaling for coordinated uplink reception (CoUR) \cite{lee12}.
An important feature of CoUR is to decouple the multiple points which
{\red send downlink signals from those which receive uplink information.}
The implementation of CoUR includes coordination and exchange of information among reception points,
computation of the receiving combiner,
{\red coordinated designation, and exchange of received data.}
The energy consumption of end users is not negligible and should be considered while implementing these techniques
and optimizing CoUR networks.
By exploiting energy harvesting, utilization, redistribution, and management policies at the user side,
energy-efficient operations of the CoMP network become promising.
New challenges of optimizing the design of CoUR include low latency,
imperfect CSI, increased signaling overhead, EH and coordination at the user side,
SE, EE and computational complexities.

\subsubsection{Dynamic CoMP Clustering}
Dynamic CoMP clustering can be increasingly complex with growing signaling overhead.
Nonetheless, it is responsive to network changes, as compared to static CoMP scenarios.
Inter-cluster interference can be minimized and the cluster size of users can be optimized adaptively and dynamically.
To optimize SE, EE, load balancing, QoS, and backhaul scheduling for dynamic clustering
in RES-powered CoMP systems,
methodologies such as greedy algorithms \cite{barron08}, game theoretic approaches \cite{song14},
and multi-stage optimizations \cite{pere} have been utilized.
However, those existing dynamic CoMP clustering approaches may lack scalability
and can suffer from a high complexity of scheduling and precoding designs.
It is a challenge to provide fully dynamic clustering at low complexities and costs when there are
increased scheduling and signaling overhead.
To this end, machine learning (ML) algorithms \cite{blum98, jiang17} can potentially help design the ahead-of-time and online resource allocation schemes of systems with the integration of unpredictable and non-dispatchable RES.
ML and Big Data techniques \cite{gupta16} can facilitate dynamic energy sharing, traffic scheduling, and CoMP clustering in a network
by considering user locations, traffic demands, latency, imperfect CSI, EH amount, and electricity prices;
and therefore deserve comprehensive investigations.

\subsection{Multi-Hop Wireless Links}
As shown in Section \ref{sec.multihop}, an interesting aspect of multi-hop networks is that network nodes can benefit from information and energy cooperation when they can share information and energy with each other.
This opens up the following potential research opportunities in this context.

\subsubsection{Opportunistic Routing}
{\red Opportunistic routing can achieve reliable data transmission for disconnected and sparse multi-hop networks,} and provide flexibility and easy adaptation to high system dynamics \cite{Shaikh2018Routing},
e.g. time-varying channels, bursty data arrivals, and intermittent EH.
The routing techniques utilize the broadcast nature of radio.
{\red Relay nodes are opportunistically chosen to forward packets.
This procedure continues until all packets are successfully delivered.}
Opportunistic routing also supports a high amount of data transfers.
Existing studies on opportunistic routing and EH are limited.

\subsubsection{Impact of Mobility on EH-Based Multi-Hop Routing}
Proper mobility models can estimate the movement of nodes with regards to position, direction, and velocity.
This can be readily captured in emerging D2D communications~\cite{Shaikh2018Routing}.
By observing the movement pattern of nodes, several works have proposed proper mobility models
and studied the impact of mobility on EH for one-hop links \cite{Niyato2014Delay,Coarasa2013Impact}.
Yet, effective resource allocation schemes that can integrate the characteristics of EH-based multi-hop routing are still missing.
The impact of mobility on the routing decisions for multi-hop networks is to be investigated.

\subsubsection{Network Coding-Aware Routing}
Network coding-aware (NC-aware) routing techniques use omni-directional antennas to utilize the broadcast nature of wireless channels. This technique shows many performance metrics over conventional routing, including improved system capacity and reliability,
and reduced delay and energy consumption \cite{Shaikh2018Routing}.
It is proven in \cite{Douik2017Instantly} that NC is effective for achieving the maximum data flow in D2D networks.
Therefore, NC-aware routing can be potentially used in multi-hop networks for performance improvement.
How to exploit RES and integrate the EH feature of multi-hop networks in NC-aware routing is an open problem.

\subsection{SWIPT Systems}
The application and integration of the SWIPT technique in SISO, MISO, MIMO, relay, and mmWave systems rely on
power allocation (transmitter) and power splitting (receiver) schemes, as unveiled in Section \ref{sec.wpt}.
The following topics on SWIPT enabled systems are worth in-depth studies.

\subsubsection{Security of SWIPT}

Increasing the transmit power can have double-sided effects on SWIPT systems.
On one hand, the desired power transferred from a
{\red source to a legitimate destination can be enhanced.}
On the other hand,  the undesired risk of information stealth by an eavesdropper can be escalated,
leading to a security concern in SWIPT systems.
{\red How to enhance the signal intensity on the legitimate recipient side while reducing the signal intensity on the eavesdropping side}
can be an exciting research direction to pursue.

\subsubsection{SWIPT for CoMP Systems}

{\red Currently, CoMP systems are categorized into two groups: joint transmission (JT)
in which an end user is served by several BSs and its data is shared globally,
versus coordinated beamforming (CB) where an end user is supported only by a single BS and its information is owned locally.
From the practicality perspective, CB-CoMP is preferred over JT-CoMP
since it requires much less signal communication overhead.}

{\red It is useful to examine the merits and challenges of incorporating SWIPT techniques into CoMP,}
where full-scale cooperations can reduce the total transmit power.
However, an enormous backhaul capacity would be needed to integrate CoMP with SWIPT
if all BSs and end users are involved in the energy transfer and information sharing.
Interference management and mitigation also needs to be dealt with in such a system.

\subsubsection{Robust Designs of SWIPT Systems}
Information transmission and EH process can have dynamic and time-varying characteristics,
rendering difficulties for the transmitters to obtain the accurate CSI ahead of time.
Offline ahead-of-time transmission policies with imprecise CSI may cause a large outage rate of the system.
It is a necessary and challenging task to design robust beamforming schemes to cope with the dynamics
of SWIPT systems.

\subsection{Energy Trading and Planning}
As discussed in Sections \ref{sec.sharing},  \ref{sec.purchasing}, and \ref{sec.trading},
the most popular optimization metrics of smart grid-powered wireless communications networks are
energy consumption and operational cost, followed by operators' utility and QoS.
The Lagrange dual based method, multi-stage optimization, and game theory are among the most
widely used techniques for offline scheduling,
while SGD and Lyapunov-based algorithms are typically applied to online optimization problems,
as collated in Table \ref{tab.trade}.
The Lagrange dual based technique provides global optimality for convex problems,
while the Lyapunov-based algorithms are asymptotically optimal for online optimizations under i.i.d. environments.
The following are some interesting research directions to be further pursued.

\subsubsection{Energy Harvesting at End Users}
Many existing works focus on EH-based BSs and the energy consumption in the downlink.
Mobile end users also need to consume energy for information reception, signal decoding, and communication in the uplink.
To fully realize self-sustained communication systems, new frameworks need to be in place to integrate the
EH capabilities for the end users in system design.
In such a system, it is important to carry out grouping among end users for
energy sharing, power allocation, interference alignment, QoS enhancement, and efficiency improvement.
The information exchange and computational complexity in the system can increase dramatically
with the growth of scheduling variables and dimensions.
Distributed optimization techniques can help alleviate the complexity of the network and protect private information of each user,
and deserve research effort.

\subsubsection{Application of Machine Learning}

ML has been one of the most active research fields due to its great success in many domains,
such as pattern recognition and computational learning theory in artificial intelligence.
It develops algorithms to learn from the past and make predictions in complicated scenarios \cite{jiang17}.
{\red ML can be widely applied to model and analyze technical problems of 5G/B5G networks.}
In particular, 5G/B5G smart end users are expected to access different spectral bands autonomously with the aid of
learning schemes of spectral efficiency.
Future Internet service providers are expected to control their transmit power and adjust their transmit protocols
with the help of QoS learning.
Future smart grid-powered BSs are expected to schedule and allocate their power according to the demand side response from the users,
and carry out online load sharing and traffic offloading based on dynamic user locations, QoS requirements, and EH amounts.


{\red Typical supervised learning methodologies depend on preexisting models and labels
that can support the extrapolation of unknown parameters
\cite{carn07, alpa14}.
They are applicable to channel estimation, data analytics, spectrum assignment, and EH amount prediction.}
They can also be applied to extrapolate locations and behaviors of mobile end users,
which can help enhance users' QoS.
{\red Unsupervised learning depends upon the input statistics in a heuristic fashion \cite{hofmann01, nieble08}.
It can be leveraged for dynamic BS or cell clustering in collaborative systems,
the association of APs in ubiquitously availing WiFi environments, and load balancing in HetNets.}
Reinforcement learning depends on a dynamic iterative learning and decision-making process \cite{kael96, sutton98}.
It may potentially be utilized to model the EH process as a Markov decision process without historical data.
It can be applied to infer the decisions made by users under unknown network conditions.
It can also facilitate the solution for the game theoretic problems between smart grid, BSs, and users.
Right now, uses of these learning techniques into the EH powered communication network design are almost an open topic.

\section{Conclusion}
We have provided a contemporary and comprehensive survey on recent breakthroughs on the utilization, redistribution, trading and planning of energy harvested in future wireless communication networks connected to smart grids.
We have reviewed a wide range of energy harvesting-based  wireless architectures such as
point-to-point, multipoint-to-point, multipoint-to-multipoint, multi-hop, and multi-cell architectures,
with an emphasis on their energy-efficient operations.
We have also gone through the SWIPT technologies as an extension of energy harvesting wireless networks.
A significant part of the article has been devoted to the redistribution redundant energy within wireless networks,
predictive planning and purchase of energy from smart grids,
and two-way trading of energy between the wireless networks and smart grids.
By redistributing and trading redundant energy, wireless service providers can reduce their electricity bills
and the consumption of brown energy.
Future research topics  on each of these different aspects of energy harvesting wireless networks
and their interoperability with smart grids have also been discussed.



\balance

\end{document}